\documentclass{aa}   	% abstract by default

\usepackage{graphicx}
\usepackage{txfonts}
\usepackage{natbib}
\bibpunct{(}{)}{;}{a}{}{,} % to follow the A&A style
			
\usepackage{graphicx}% Include figure files
\usepackage{dcolumn}% Align table columns on decimal point
\usepackage{bm}% bold math
\usepackage{journaux}

\newcommand{\fig}{\textsc{Figure~}}

\newcommand{\bB}{\bm{B}}

\newcommand{\FLor}{\bm{F}_{\!\mathbf{{\mathcal{L}}}}}
\newcommand{\bnab}{\bm{\nabla}}

\newcommand{\er}{\mathbf{\hat{e}_{r}}}
\newcommand{\etheta}{\mathbf{\hat{e}_{\theta}}}
\newcommand{\ephi}{\mathbf{\hat{e}_{\varphi}}}
\newcommand{\mbf}[1]{\textbf{\textit{#1}}}

\begin{document}
%-----------------------------------------------------------------------

\title{Relaxed Equilibrium Configurations to Model Fossil Fields\\ I -- A first family}

%--- Authors --------------------------------------------------------

   \author{V. Duez 
     %\inst{1}
          \and
          S. Mathis
          %\inst{1}\fnmsep
          }

   \institute{CEA/DSM/IRFU/SAp, CE Saclay, F-91191 Gif-sur-Yvette Cedex, France;
   		%\\ \noindent 
		AIM, UMR 7158, CEA - CNRS - Universit\'e Paris 7, France\\
              \email{vincent.duez@cea.fr,stephane.mathis@cea.fr}
             }

   \date{Received 19 October 2009 / Accepted 06 March 2010} %hypothetical !

%-----------------------------------------------------------------------
  \abstract
  % Context heading (optional) 
{The understanding of fossil fields origin, topology and stability is one of the corner stones of the stellar magnetism theory. On one hand, since they survive over secular time-scales, they may modify the structure and the evolution of their host stars. On the other hand, they must have a complex stable structure since it has been demonstrated by Tayler and collaborators that simplest purely poloidal or toroidal fields are unstable on dynamical time-scales. In this context, the only stable configuration which has been found today is the one resulting of a numerical simulation by Braithwaite and collaborators who have studied the evolution of an initial stochastic magnetic field, which is found to relax on a mixed stable configuration (poloidal and toroidal) that seems to be in equilibrium and then diffuses.}
  % Aims heading (mandatory)
   {In this work, we thus go on the track of such type of equilibrium field in a semi-analytical way.}
  % Methods heading (mandatory)
   {In this first article, we study the barotropic magnetohydrostatic equilibrium states; the problem reduces to a Grad-Shafranov-like equation with arbitrary functions. Those latters are constrained by deriving the lowest-energy equilibrium states for given invariants of the considered axisymmetric problem and in particular for a given helicity which is known to be one of the main actor of such problems. Then, we obtain the generalization of the force-free Taylor's relaxation states obtained in laboratory experiments (in spheromaks) that become non force-free in the self-gravitating stellar case.  The case of general baroclinic equilibrium states will be studied in Paper II.}
  % Results heading (mandatory)
   {Those theoretical results are applied to realistic stellar cases, namely to the solar radiative core and to the envelope of an Ap star, and discussed. In both cases we assume that the field is initially confined in the stellar radiation zone. %delimited by the surrounding convective material.
   }
  % Conclusions heading (optional)
  {}
%-----------------------------------------------------------------------
%-----------------------------------------------------------------------

\titlerunning{Relaxed Equilibrium Configurations to Model Fossil Fields}
\authorrunning{V. Duez \& S. Mathis}
\maketitle

%
%\newpage
%****************************************************************************
%**************************** Introduction **********************************
%****************************************************************************
\section{\label{intro} Introduction}
Spectropolarimetry is nowadays exploring the stellar magnetism across the whole Hertzsprung-Russel diagram 
\citep{Donati:1997, Donati:2006, Neiner:2007, Landstreet:2008, Petit:2008}.
%(Donati et al. 1997, Donati et al. 2006, Neiner 2007, Landstreet et al. 2008, Petit et al. 2008). 
Furthermore, helioseismology and asteroseismology are providing new constraints on internal transport processes occuring in stellar interiors 
\citep{Turck-Chieze:2008, Aerts:2008}. 
%(Turck-Chi\`eze \& Talon 2008, Aerts et al. 2008). 
In this context, even if standard stellar models explain the main features of stellar evolution, it is now crucial to go beyond this modelling to introduce dynamical processes such as magnetic field and rotation to investigate their effects on stellar structure and secular evolution 
\citep{Maeder:2000, Talon:2008}. 
%(Maeder \& Meynet 2000, Talon 2008). 
To achieve this aim, secular MHD transport equations have been derived in order to be introduced in stellar evolution codes. They take into account in a coherent way the interaction between differential rotation, turbulence, meridional circulation, and magnetic field 
\citep{Spruit:2002, Maeder:2004, Mathis:2005}, 
%(Spruit 2002, Maeder \& Meynet 2004, Mathis \& Zahn 2005), 
while non-linear numerical simulations provide new insight on these mechanisms 
\citep{Charbonneau:1993, Rudiger:1997, Garaud:2002, BrunZahn:2006}. 
%(Charbonneau \& MacGregor 1993, R\"udiger \& Kitchatinov 1997, Garaud 2002, Brun \& Zahn 2006). 
If we want to go further, the simplest modifications of static structural properties such as density, gravity, pressure, temperature, and luminosity induced by the magnetic field have also to be systematically quantified as a function of the field geometry and strength 
\citep{Moss:1973, Mestel:1977, Lydon:1995, Couvidat:2003, Li:2006, Duez:2008, Li:2009, Duezetal:2010}. 
%(Moss 1973, Mestel \& Moss 1977, Lydon \& Sofia 1995, Couvidat et al. 2003, Li et al. 2006, Duez et al. 2008, Li et al. 2008). 

However, an infinity of possible magnetic configurations can be investigated since the different observation techniques only lead to indirect indications on the internal field topologies through the surface field properties they provide. Furthermore, since the simplest geometrical configurations like purely poloidal and purely toroidal fields are known to be unstable 
\citep{Acheson:1978, Tayler:1973, Markey:1973, Markey:1974, Goossens:1978, Goossens:1980, Goossens:1981, van-Assche:1982, Spruit:1999, Braithwaite:2006b, Braithwaite:2007}, 
%(Achesson 1978, Tayler 1973, Markey \& Tayler 1973-1974, Goossens \& Veugelen 1978, Goossens \& Tayler 1980, Goossens, Biront \& Tayler 1981, Van Assche et al. 1982, Spruit 1999, Braithwaite 2006b-2007), 
the best candidates for stable geometries are mixed poloidal-toroidal fields 
\citep{Wright:1973, Markey:1974, Tayler:1980, Braithwaite:2009}. 
%(Wright 1973, Markey \& Tayler 1974, Tayler 1980, Braithwaite 2008). 

Therefore, it is necessary to go on the track of possible stable magnetic configurations in stellar interiors to evaluate their effects on stellar structure and to use them as potential initial conditions to study secular internal transport processes.\\

{In this work, we thus revisit the pioneer works by 
\cite{Ferraro:1954, Mestel:1956, Prendergast:1956} and \cite{Woltjer:1960}. 
\cite{Ferraro:1954} studied the equilibrium configurations of an incompressible star with a purely poloidal field. \cite{Prendergast:1956} \citep[see also][]{Chandrasekhar:1956a, Chandrasekhar:1956b,Chandrasekhar:1956c} then extended the model to take into account the toroidal field, by solving the magneto-hydrostatic equilibrium of incompressible spheres.} The obtained configurations  seem to be relevant in regard of the most recent numerical simulations that may explain fossil fields in early-type stars, white dwarfs or neutron stars 
\citep{Braithwaite:2004, Braithwaite:2006, Braithwaite:2006b} 
%(cf. Braithwaite \& Spruit 2004, Braithwaite \& Nordlund 2006, Braithwaite \& Spruit 2006) 
and of theoretical studies of their helicity relaxation 
\citep{Broderick:2008, Mastrano:2008}.  
%(Broderick \& Narayan 2008, Mastrano \& Melatos 2008). 
The main generalization of Prendergast's work we achieve here consists in relaxing the incompressible hypothesis in order to take into account the star's structure \citep[see also][]{Woltjer:1960}, which differs as a function of its stellar type and of its evolution stage, and to derive the minimum energy equilibrium configuration for a given mass and helicity, which are then applied to realistic models of stellar interiors. 

Assuming that the Lorentz volumetric force is a perturbation compared with the gravity, we derive the non force-free magnetohydrostatic equilibrium. In this first article, we then focus on the barotropic equilibrium states family \footnote{Barotropic states are such that their density and pressure gradients are aligned. They can be convectively stable or not, depending on their entropy stratification. We will precisely introduce their definition in \S 2.2. 
%A spherical standard stellar model (without rotation nor magnetic field) is in a barotropic state in its non-convective regions.
 }, for which the possible field configurations and the stellar structure are explicitely coupled; these may correspond to the numerical experiments by Braithwaite and collaborators. In this case, the problem reduces to a Grad-Shafranov-like equation 
\citep{Grad:1958, Shafranov:1966, Kutvitskii:1994}, 
%(Grad \& Rubin 1958, Shafranov 1966, Kutvitskii \& Solov'ev 1994), 
similar to the one intensively used in fusion plasma physics. We then focus on its {minimum energy} eigenmodes {for a given mass and helicity}, which are derived and applied in order to model {relaxed} stellar fossil magnetic fields which are found to be non force-free. {Arguments in favor of the stability of the obtained configurations are finally discussed \citep{Wright:1973, Tayler:1980, Braithwaite:2009, Reisenegger:2009} and we compare their properties with those of relaxed fields obtained in numerical simulations \citep{Braithwaite:2008M}}. The case of general baroclinic equilibrium states will be studied in Paper II \citep{Wright:1969,Moss:1975}.
%(Wright 1973, Tayler 1980, Braithwaite 2008, Reisenegger 2008). 
%Finally, we discuss the baroclinicity, which will naturally takes place inside the star because of the interaction of the field with differential rotation, turbulence and meridional circulation.
%
%****************************************************************************
%***************     Section : The Non Force-Free MHS Equilibrium ****** 
%****************************************************************************
\section{\label{defs} The Non Force-Free Magneto-Hydrostatic Equilibrium}
In this work, we focus on the magnetic equilibrium of a self-gravitating spherical shell to model fossil fields in stellar interiors. To achieve this goal, we start from
\begin{equation}
\mbf 0=-\bnab P-\rho\bnab V+\FLor,\, \hbox{where} \,\, \FLor=\mbf j \times \mbf B.
\label{MHS1}
\end{equation}
Eq. (\ref{MHS1}) must be satisfied in the interior of an infinitely conducting mass of fluid in the presence of a large-scale field
together with the Poisson equation, $
\nabla^{2}V=4\:\pi\: G\:\rho$, and the Maxwell equations,
$\bnab\cdot\mbf B=0$ (Maxwell flux) and 
$\bnab\:\mathbf{\times}\:\mbf B=\mu_0\:\:\mbf j$ (Maxwell-Amp\`ere).\\
$P$, $\rho$ and $V$ are respectively the pressure, the density and the gravitational potential of the considered plasma. $\mbf B$ is the magnetic field and $\mbf j$ is the associated current, which is given in the classical MHD approximation by the Maxwell-Amp\`ere's equation. $\mu_0$ is the magnetic permeability of the plasma and $\FLor$ is the Lorentz force.
%
%\subsection{Magnetic field configuration and the magnetohydrostatic equilibrium}
\subsection{\label{axiB} Magnetic field configuration and the magnetohydrostatic equilibrium}
If we consider only the axisymmetric case, where all physical variables are independent of the azimuthal angle ($\varphi$), $\mbf B\left(r,\theta\right)$ can be written in the form
\begin{equation}
\mbf B=
\frac {1}{r \sin \theta} \bnab \Psi\left(r,\theta\right) \mathbf{\times}\, \ephi + \frac{1}{r \sin \theta}\, F\left(r,\theta\right) \,\ephi\,,
\label{BPsiF}
\end{equation}
which is divergenceless. $\Psi$ and $F$ are respectively the poloidal flux function and the toroidal potential; $\left(r,\theta,\varphi\right)$ are the usual spherical coordinates and $\left\{{\bf {\hat e}}_{k}\right\}_{k=r,\theta,\varphi}$ their unit-vector basis. Finally, the poloidal component of the magnetic field ($\mbf B_{\rm P}$) is such that $\mbf B_{\rm P}\cdot\bnab\Psi=0$, so it belongs to iso-$\Psi$ surfaces.
The magnetohydrostatic equilibrium (Eq. \ref{MHS1}) implies that the poloidal part (in the meridional plane in the axisymmetric case) of the Lorentz force ($\textit{\textbf{F}}_{\!\mathbf{{\mathcal L}_P}}$) balances the pressure gradient and the {gravitational} force, which are also purely poloidal vectors, while, in the absence of any other force, its toroidal component ($\textit{\textbf{F}}_{\!\mathbf{{\mathcal L}_T}}=\textit{F}_{\!\mathbf{{\mathcal L}_{\varphi}}}\bf{\hat e}_{\varphi}$) vanishes.
We thus have
\begin{equation}
\FLor=\textit{\textbf{F}}_{\!\mathbf{{\mathcal L}_P}}+\textit{F}_{\!\mathbf{{\mathcal L}_{\varphi}}}\bf{\hat e}_{\varphi}=\textit{\textbf{F}}_{\!\mathbf{{\mathcal L}_P}}.
\end{equation}
Using Eq. (\ref{BPsiF}), we obtain
\begin{eqnarray}
\textit{\textbf{F}}_{\!\mathbf{{\mathcal L}_P}}=-\frac{1}{\mu_0 r^2 \sin^2 \theta}\Bigg\{\left(F\partial_r F+\partial_r\Psi\Delta^{*}\Psi\right){\bf\hat e}_{r}\nonumber\\
{+\frac{1}{r}\left(F\partial_{\theta} F+\partial_{\theta}\Psi\Delta^{*}\Psi\right){\bf\hat e}_{\theta}\Bigg\}},
\end{eqnarray}
with $\partial_{x}=\partial/\partial x$, and
\begin{equation}
\Delta^* \Psi \equiv \partial_{rr}{\Psi} + \frac{\sin \theta}{r^2}\partial_{\theta}\left(\frac{1}{\sin \theta}\partial_{\theta}{\Psi}\right)
\label{gsoperator}
\end{equation}
is the usual Grad-Shafranov operator in spherical coordinates.
On the other hand, since $\textit{F}_{\!\mathbf{{\mathcal L}_{\varphi}}}=0$, we get
$\partial_r\Psi\partial_{\theta}F-\partial_\theta\Psi\partial_r F= 0$;
the non-trivial values for $F$ are therefore obtained by setting
\begin{equation}
F(r, \theta) = F(\Psi).
\label{FPsi}
\end{equation}
Then, we obtain
$F\partial_{r}{F}=F\partial_{\Psi}F\partial_r\Psi$ and $F\partial_{\theta}{F}=F\partial_{\Psi}F\partial_{\theta}\Psi$, leading to the final expansion of the Lorentz force
\begin{equation}
\textit{\textbf{F}}_{\!\mathbf{{\mathcal{L}}}} = {\mathcal A}\left(r,\theta\right) \bnab \Psi,
\label{LorentzPsi}
\end{equation}
where
\begin{equation}
{\mathcal A}(r, \theta) = -\frac{1}{\mu_0 r^2 \sin^2 \theta} \left(  F\partial_{\Psi} F +  \Delta^{*} \Psi \right).
\label{Aux}
\end{equation}
Therefore, the poloidal component of $\textit{\textbf{F}}_{\!\mathbf{{\mathcal{L}}}}$ is nonzero a priori, the field being thus non force-free in this case.

{This point has here to be discussed. First, \cite{Reisenegger:2009} demonstrated that the magnetic field can not be force-free everywhere in stellar interiors (see the demonstration in the appendix A of his paper). Note in this context that the ``force-free'' configurations obtained by \cite{Broderick:2008} verify this theorem because they have current sheets with a non-zero Lorentz force on the stellar surface. Moreover, \cite{Shulyaketal:2007,Shulyaketal:2009} showed how the atmosphere of a CP star can be the host of a non-zero Lorentz force. Therefore, from now on, we consider the non force-free equilibrium.}\\
%, so the pressure gradient, the density, and the gravific potential adjust in order to relax towards equilibrium ruled by Eq. (\ref{MHS1}).
%In contrast, the field is force-free in the azimuthal direction by construction.
%
%\subsection{The barotropic equilibrium states family}
%\subsection{The barotropic equilibrium states family}

Now, if we take the curl of Eq. (\ref{MHS1}), we get the static vorticity equation
\begin{equation}
-\frac{\bnab\,\rho\times\bnab\,P}{\rho^2}=\bnab\times\left(\frac{\textit{\textbf{F}}_{\!\mathbf{{\mathcal{L}}}}}{\rho}\right),
\label{curlmhs}
\end{equation}
that governs the balance between the baroclinic torque (left-hand side; see 
\cite{Rieutord:2006} 
%Rieutord 2006 
for a detailed description) and the magnetic source term. Then, as has been emphasized by \cite{Mestel:1956}, 
the different structural quantities such as the density, the {gravitational} potential, and the pressure relax in order to verify Eq. (\ref{MHS1}) for a given field configuration 
(see \cite{Sweet:1950, Moss:1975}, and \cite{Mathis:2005} \S 5.). 
%(see Sweet 1950 and Mathis \& Zahn 2005 \S 5.). 
Thus, the choice for $\Psi$ is left free.\\ 
%{ Let us emphasize that \cite{Mestel:1956} argued that the strict barotropic condition cannot be completely fulfilled in the general case. Recently, this has been discussed by \cite{Haskell:2008}, who gave the generalization of the zero-torque equation above, in the case of a general equation of state. However, we have to underline that the vanishing of the above equation can also be obtained for a non-barotropic equation of state, under the condition that the fluid is barotropic (in the hydrodynamic acceptance of the term). As a matter of fact, for an equation of state on the form $\rho = \rho(p,T)$ {\it \bf e. g.}, the only condition that need to be fulfilled is that the pressure, the temperature and the density gradient are aligned: this does not require a formal dependence of two quantities on the third one only. Any model computed with a one-dimensionnal stellar evolution code embodies this statement.\\
%As done by \cite{Haskell:2008}, we will here restrict to the simplest case in which it is assumed that the magnetic field will only produce small deviations from a spherically symmetric background model.
%}

\subsection{The barotropic equilibrium state family}

Magnetic initial configurations are one of the crucial unanswered question for the modelling of MHD transport processes in stellar interiors. To examine this question, Braithwaite and collaborators 
\citep{Braithwaite:2004, Braithwaite:2006} 
%(Braithwaite \& Spruit 2004, Braithwaite \& Nordlund 2006)
 studied the relaxation of an initially stochastic field in models of convectively stable stellar radiation zones. The field is found to relax, after several Alfv\'en times, to a mixed poloidal-toroidal equilibrium configuration, which then diffuses towards the exterior.\\ 

We choose here, using an analytical approach, to find such field geometries, which are governed at the beginning by the magnetohydrostatic equilibrium. 

To achieve this aim, we focus in this first article on the particular barotropic equilibrium states (in the hydrodynamic {meaning} of the term) for which the field configuration is explicitely coupled with the stellar structure, since we have in this case
\begin{equation}
-\frac{\bnab\,\rho\times\bnab\,P}{\rho^2}=\bnab\times\left(\frac{\textit{\textbf{F}}_{\!\mathbf{{\mathcal{L}}}}}{\rho}\right)=\mbf 0.
\label{smhs}
\end{equation}
Those are the generalization of the Prendergast's equilibria which take into account the compressibility and which have been studied in polytropic cases by
\cite{Woltjer:1960, Wentzel:1961, Roxburgh:1966, Monaghan:1976}. 
%We will now examine them using realistic 1D stellar models.

Let us {first} recall the definition of the barotropic states. In fluid mechanics, a fluid is said to be in a barotropic state if the following condition is satisfied (see \cite{Pedlosky:1998} in a geophysical context and \cite{Zahn:1992} in a stellar one):
\begin{equation}
\bnab\rho\times\bnab P={\mbf 0};
\label{Eq1}
\end{equation}
in other words, the baroclinic torque in the vorticity equation (Eq. \ref{curlmhs}) vanishes. Then, the surfaces of equal density coincide with the isobars since the density and the pressure gradients are aligned. This does not imply any question of equation of state, which in stellar interiors can take the most general form $P=f\left(\rho,T,\cdot\!\cdot\!\cdot\,\right)$ ($T$ being the temperature). Moreover, this does not presume anything about the stratification of the fluid, which can be stably stratified or not. For example, a star in solid body rotation is in a barotropic state (as opposed to baroclinic) \citep[see once again][]{Zahn:1992}. Let us illustrate this point with the simplest case of a non-rotating and non-magnetic stellar radiation zone.
In this case, the hydrostatic balance is given by
$
{\bnab P}/{\rho}=-\bnab V={\mbf g}
$. If we take the curl of this equation, we obtain the stationary version of the thermal-wind equation:
\begin{equation}
-\frac{\bnab\rho\times\bnab P}{\rho^2}=-\bnab\times\left[\bnab V\right]={\mbf 0}
\end{equation}
and the star is thus in a barotropic state in the hydrodynamic {meaning} of the term.

This has to be distinguished from the point of view of thermodynamics where a barotropic equation of state is such that
$P=f\left(\rho\right)$ while a non-barotropic equation of state is such that $P=f\left(\rho,T,\cdot\!\cdot\!\cdot\,\right)$. 

Then, it is clear that a fluid with a barotropic equation of state is automatically in an hydrodynamical barotropic state. However, in the case of a fluid with a non-barotropic equation of state, the situation is more subtle:  in the case where the curl of the volumetric perturbing force vanishes ({i.e.} $\bnab\times\left({\textit{\textbf{F}}_{\!\mathbf{{\mathcal{L}}}}}/{\rho}\right)=0$) the fluid is in an hydrodynamical barotropic state, while in the general case it is in a baroclinic situation. Then, a fluid with a non-barotropic equation of state can be in a barotropic state even if it is only for a specific form of the perturbing force. In this first work, we choose to examine the first equilibrium family in which the Lorentz force verify the barotropic balance described by Eq. (\ref{Eq1}) in a stably stratified radiation zone. The second general case ({cf.} Mestel 1956) will be studied in paper II.\\

%Braithwaite and collaborators approximate stable stellar radiation zones by a polytrope with polytropic index $n=1/\left(\Gamma-1\right)\approx3$ to study the relaxation of fossil magnetic fields and find the only known stable equilibrium configuration which is thus a barotropic equilibrium state which verify Eq. (\ref{Eq1}) due to Eq. (\ref{Eq2}-\ref{Eq3}). This is the reason why we choose to study the specific barotropic equilibria family which verify Eq. (\ref{Eq1}).\\
%The case we considered here is thus that of a stably stratified radiation zone, in which the Lorentz force is chosen such as to verify the barotropic balance described by Eq. (\ref{Eq1}).\\ 

Stellar interiors, except just under the surface, are in a regime where $\beta=P/P_{\rm Mag}\!>\!\!>\!1$, $P_{\rm Mag}=B^2/(2\,\mu_0)$ being the plasma's magnetic pressure. On the other hand, in the domain of fields amplitudes relevant for classical stars (i.e. the non-compact objects), the ratio of the volumetric Lorentz force %(which is the sum of the gradient of the magnetic pressure and of the magnetic tension force)
by the gravity is very weak. Therefore, the stellar structure modifications induced by the field can be considered as perturbations only from a spherically symmetric background \citep{Haskell:2008}. Then, we can write $\rho\approx\overline\rho+\widetilde\rho$, where $\overline\rho$ and $\widetilde\rho$ are respectively the mean density on an isobar, which is given at the first order by the standard non-magnetic radial density profile of the considered star, and its magnetic-induced perturbation on the isobar (with $\widetilde\rho\!<\!\!<\!\overline\rho$). %Their evolution will be mainly modified by transport processes due to magnetic field % but not by its direct impact on the structure
% and to the direct impact of the Lorentz force on the structure just below the stellar surface where the strong $\beta$ regime breaks down. 
Thus to the first order, Eq. (\ref{smhs}) on an isobar becomes \begin{equation}-\frac{\bnab \widetilde\rho\times\mbf g_{\rm eff}}{\overline\rho}=\bnab\times\left(\frac{\textit{\textbf{F}}_{\!\mathbf{{\mathcal{L}}}}}{{\overline\rho}}\right)=\mbf 0,\end{equation} where the effective gravity $({\mbf g}_{\rm eff})$, such that $\bnab\overline P=\overline\rho\:{\mbf g}_{\rm eff}$, has been introduced.
%\begin{eqnarray}
%-\frac{\bnab\,\rho\times\bnab\,P}{\rho^2}\approx\bnab\times\frac{\textit{\textbf{F}}_{\!\mathbf{{\mathcal{L}}}}}{\overline\rho}%\approx-\frac{\bnab\,\overline\rho\times\bnab\,\overline P}{\overline\rho^2} 
%\approx\mathbf 0
%\label{curlmhsf}
%\end{eqnarray}
This gives, using Eq. (\ref{LorentzPsi})
\begin{eqnarray}
\bnab\left(\frac{\mathcal A}{\overline\rho}\right) \times \bnab \Psi = {\mbf 0},
\label{curlFL}
\end{eqnarray}
which projects only along $\ephi$ as
\begin{eqnarray}
\partial_{r}\left(\frac{\mathcal A}{\overline\rho}\right)\partial_{\theta}\Psi- \partial_{\theta}\left(\frac{\mathcal A}{\overline\rho}\right)\partial_{r}\Psi= 0,
\end{eqnarray}
so that there exists a function $G$ of $\Psi$ such that
\begin{eqnarray}
\frac{\mathcal A}{\overline\rho}=G\left(\Psi\right).
\label{G}
\end{eqnarray}
Then, Eq. (\ref{Aux}) leads to the following one ruling $\Psi$
\begin{equation}
\Delta^{*}\Psi+F\left(\Psi\right)\partial_{\Psi}\left[F\left(\Psi\right)\right]=-\mu_0 r^2 \sin^2\theta\,\overline\rho\, G\left(\Psi\right).
\label{GS}
\end{equation}
This equation is similar to the well-known Grad-Shafranov equation\footnote{The usual Grad-Shafranov equation is given by:\\
$\Delta^{*}\Psi+F\left(\Psi\right)\partial_{\Psi}\left[F\left(\Psi\right)\right]=-\mu_{0}r^2\sin^2\theta\partial_{\Psi}\left[P\left(\Psi\right)\right]
$, \\
where the pressure $P$ is prescribed in function of $\Psi$. These describes the equilibrium between the magnetic force and the pressure gradient only.} which is used to find equilibria in magnetically confined plasmas such as those in tokamaks or in spheromaks 
\citep{Grad:1958, Shafranov:1966}.
%(Grad \& Rubin 1958, Shafranov 1966). 
However, here the source term is different and is directly related to the internal structure of the star through its density profile ($\overline\rho$) {(the general form of the Grad-Shafranov equation in an astrophysical context is discussed for example in \cite{HeinemannOlbert:1978} and \cite{Ogilvie:1997})}. {Moreover, since the field has to be non force-free in stellar interiors $G\ne0$ (see the previous discussion in \S 2.1. and Eqs. \ref{G}, \ref{LorentzPsi} and \ref{Aux}).}

It is only applicable to the case of the barotropic state family. The equations for the general case will be studied in paper II.

\subsection{F and G expansion}

Let us now focus on the respective expansion of $F$ and $G$ as a function of $\Psi$. 

First, since $F$ is a regular function, we can expand it in {power series in} $\Psi$:
\begin{equation}
F(\Psi) = \sum_{i=0}^{\infty}\frac{\lambda_{i}}{R} \Psi^{i},
\end{equation}
the $\lambda_i$ being the expansion coefficients that have to be determined and $R$ a characteristic radius which will be identified below. On the other hand, $B_{\varphi}$ must be regular at the center of the sphere; the first term ($i=0$) of the previous expansion is then excluded (cf. Eq. \ref{BPsiF}), the above expansion thus reducing to
$F(\Psi) = \sum_{i>0}\left(\lambda_{i}/R\right)\Psi^{i}$.

{In the same way, $G$ can be expanded as}
\begin{equation}
G\left(\Psi\right)=\sum_{j=0}^{\infty}\beta_{j} \Psi^{j}.
\end{equation}
Then, Eq. (\ref{GS}) becomes:
\begin{eqnarray}
\Delta^{*}\Psi+\sum_{k>0}\frac{\Lambda_{k}}{R^2}\Psi^{k}=-\mu_0 r^2 \sin^2\theta\,\overline\rho\, \sum_{j=0}^{\infty}\beta_{j} \Psi^{j}, 
\label{GSNL}
\end{eqnarray}
where $\Lambda_k=\sum_{i_1>0}\sum_{i_2>0}\left\{i_2 \lambda_{i_1}\lambda_{i_2}\delta_{i_1+i_2-1,k}\right\}$, $\delta$ being the usual Kronecker symbol. 
This is the generalization of the Grad-Shafranov-type equation obtained by Prendergast (1956) for the barotropic compressible states.\\
%{\it Since this equation is highly non-linear, we choose here to study, as a first step, its linear eigenmodes for which variables do separate (in $r$ and $\theta$) and for which Eq. (\ref{GSNL}) can thus be solved semi-analytically; its complete non-linear resolution will be undertaken in a forthcoming work.}\\

{Thus, having assumed the {non force-free} barotropic magneto-hydrostatic equilibrium state leads to undetermined arbitrary functions ($F$ and $G$) that must be constrained. To achieve this aim, we follow the method given in the axisymmetric case by \cite{Chandrasekhar:1958} and \cite{Woltjer:1959b} that allows to find the equilibrium state of lowest energy compatible with the constancy of given invariants for the studied axisymmetric system.}
%
%****************************************************************************
%***************     Section : Self-Gravitating Relaxation States *********** 
%****************************************************************************
\section{Self-Gravitating Relaxation States}
\subsection{Definitions and axisymmetric invariants}
{We first introduce the cylindrical coordinates $\left(s,\varphi,z\right)$ where $s=r\sin\theta$ and $z=r\cos\theta$. Then, $\mbf B$ given in Eq. (\ref{BPsiF}) becomes
\begin{equation}
\mbf B\left(s,z\right)=\frac{1}{s}\bnab \Psi\left(s,z\right)\times\,\ephi+\frac{1}{s}F\left(s,z\right)\,\ephi.
\label{Cy1}
\end{equation}
Then, we define the potential vector $\mbf A\left(s,z\right)=A_{\varphi}\left(s,z\right)\ephi$ such that $\mbf B_{\rm P}=\bnab \times\mbf A$ and we get
\begin{equation}
\mbf B=\bnab \times{\mbf A}+\frac{F}{s}\ephi\quad\hbox{where}\quad A_{\varphi}\left(s,z\right)=\frac{\Psi}{s}.
\end{equation}
Next, we insert the expansion for the magnetic field $\mbf B$ used by \cite{Chandrasekhar:1958} and \cite{Woltjer:1959a,Woltjer:1959b}:
\begin{equation}
\mbf B=-s\:\partial_{z}{\Phi}\left(s,z\right)\mathbf{\hat{e}_{s}} +\!\frac{1}{s}\partial_{s}\left[s^2{\Phi}\left(s,z\right)\right]\mathbf{\hat{e}_{z}}+s \:{\mathcal T}\left(s,z\right)\ephi
\label{Cy2}
\end{equation}
where $\left\{\widehat{\mbf e}_{k}\right\}_{k=s,\varphi,z}$ is the cylindrical unit-vector basis and where we identify using Eq. (\ref{Cy1})
\begin{equation}
\Psi=s^2 {\Phi}\quad\hbox{and}\quad F=s^2 {\mathcal T}.
\label{Phitau1}
\end{equation}
The Grad-Shafranov operator applied to $\Psi$ can then be expressed as follow
\begin{equation}
\Delta^{*}\Psi=s^2\: \bnab \cdot\left(\frac{\bnab \Psi}{s^2}\right)=\left[\partial_{ss}\!-\!\frac{1}{s}\partial_s\!+\!\partial_{zz}\right]\Psi=s^2\Delta_5 \Phi,
\label{Phitau2}
\end{equation}
where
$ \Delta_{5}=\partial_{ss}+\frac{3}{s}\partial_s+\partial_{zz}. $

We now introduce the two general {families} of invariants of the barotropic axisymmetric magneto-hydrostatic equilibrium states, which have been introduced by \cite{Woltjer:1959b} for the compressible case \citep[see also][]{Wentzel:1960}:
\begin{equation}
{\mathcal I}_{{\rm I};n}=\int_{\mathcal V}M_{n}\left(s^2 \Phi\right){\overline\rho}\,{\rm d}{\mathcal V}=\int_{\mathcal V}\left(s^2 \Phi\right)^{n}{\overline\rho}\,{\rm d}{\mathcal V},
\label{I1}
\end{equation}
\vskip -5pt
\begin{equation}
{\mathcal I}_{{\rm I\!I};q}=\int_{\mathcal V}N_{q}\left(s^2 \Phi\right)\,{\mathcal T}\,{\rm d}{\mathcal V}=\int_{\mathcal V}\left(s^2 \Phi\right)^{q}{\mathcal T}\,{\rm d}{\mathcal V},
\end{equation}
where $M_{n}$ and $N_{q}$ are arbitrary functions that have to be specified. Those latters are conserved as long as
\begin{equation}
\mbf B\cdot\widehat{\mbf e}_{r}=0\quad\hbox{(\it i.e. }\Phi={\mathcal T}=0)
\end{equation}
on the boundaries. 
\begin{figure}[h!]
\begin{center}
\includegraphics[width=0.325\textwidth]{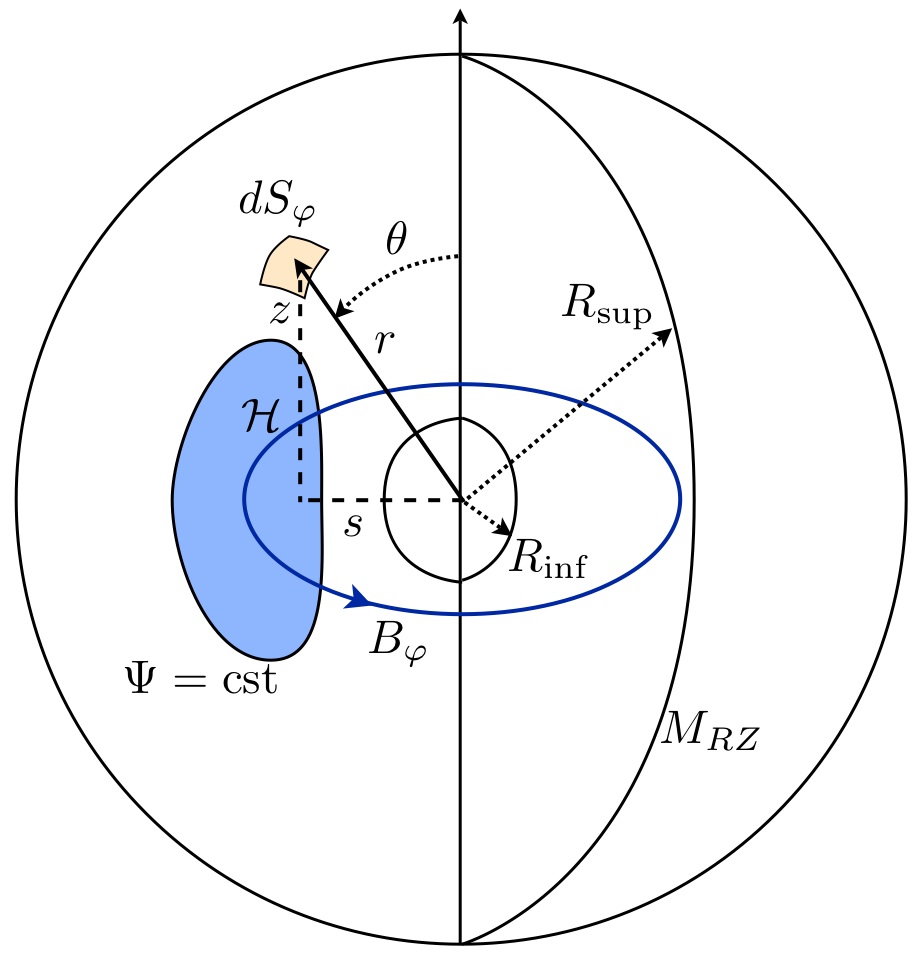}
\end{center}
%\vskip -15pt
\caption{Schematic representation of the two coordinates systems used and of a constant $\Psi$ surface. The invariants of the axisymmetric system are the total mass of the considered stellar radiative region ($M_{\rm RZ}$), the mass {enclosed} in a constant $\Psi$ surface, the toroidal flux ($\mathcal{F}_{\varphi}$) associated with the toroidal magnetic field ($B_{\varphi}$), and the global helicity ($\mathcal{H}$).}
\end{figure}

\subsection{Fossil fields barotropic relaxation states}
Let us first concentrate on ${\mathcal I}_{{\rm I\!I};q}$ and $N_{q}$ relevant for fossil fields relaxation. First, if we set $N_{0}\left(s^2 \Phi\right)=1$, we obtain
\begin{eqnarray}
{\mathcal I}_{{\rm I\!I};0}&=&\int_{\mathcal V}{\mathcal T}\,{\rm d}{\mathcal V}=2\,\pi \int_{\mathcal S} B_{\varphi}{\rm d}s{\rm d}z\nonumber\\
&=&2\,\pi \int_{\mathcal S} B_{\varphi}\:{\rm d}{\mathcal S}_{\varphi}=2\,\pi\:{\mathcal F}_{\varphi}
\end{eqnarray}
that corresponds to the conservation of the flux of the azimuthal field across the meridional plane of the star (${\mathcal F}_{\varphi}$) in perfect axisymmetric MHD equilibria.\\% (note that this has been first isolated in the incompressible case by Chandrasekhar (1958) who called it $I_6$).\\

Then, if we set $N_{1}\left(s^2\, \Phi\right)=s^2\, \Phi$, we get
\begin{equation}
{\mathcal I}_{\rm I\!I;1}=\int_{\mathcal V}\left(s^2\, \Phi\right){\mathcal T}\,{\rm d}{\mathcal V}=\int_{\mathcal V}A_{\varphi}B_{\varphi}{\rm d}{\mathcal V}%\nonumber\\
%&=&\int_{\mathcal V}\mbf A\cdot\mbf B{\rm d}{\mathcal V}
={\mathcal H}/2,
\end{equation}
where we thus identify the magnetic helicity (${\mathcal H}$; see \S 5.1.) of the field configuration which is a global quantity integrated over the volume of the studied radiation zone.
 
Let us briefly discuss the peculiar role of this quantity in the search of stable equilibria. As emphasized by \cite{Spruit:2008}, the magnetic helicity is a conserved quantity in a perfectly conducting fluid with fixed boundary conditions. However, in realistic conditions, rapid reconnection can take place even at very high conductivity, especially when the field is dynamically evolving, for example during its initial relaxation phase. Nevertheless, in laboratory experiments, like for example in spheromaks, the helicity is often observed to be approximately conserved, which leads to the existence of stable equilibrium configurations. In fact, if the helicity is conserved a dynamical or unstable field with a finite initial helicity (${\mathcal H}_{0}$) cannot decay completely, the helicity of vanishing field being zero. This is precisely what has been observed in the numerical experiment performed by \cite{Braithwaite:2004} and \cite{Braithwaite:2006}, where an initial stochastic field with a finite helicity decays initially but relaxes into a stable equilibrium. 

In the context of laboratory low-$\beta$ plasmas, this process has been identified by \cite{Taylor:1974} and is thus called the Taylor's relaxation.\\

For this reason, we now search as \cite{Chandrasekhar:1958} the {\it final state of equilibrium}, which is the state of lowest energy that the compressible star, preserving its axisymmetry, can attain while conserving the invariants ${\mathcal I}_{{\rm I};n}$, ${\mathcal I}_{\rm I\!I;0}={\mathcal F}_{\varphi}$ and ${\mathcal I}_{\rm I\!I;1}={\mathcal H}/2$ in barotropic states.\\

To achieve this, we thus introduce the total energy of the system
\begin{eqnarray}
E&=&\frac{1}{2}\int_{\mathcal V}\left\{\frac{{\mbf B}^2}{\mu_0}+{%\overline
\rho}\left[V+2{\mathcal U}\right]\right\}{\rm d}{\mathcal V}\nonumber\\
&=&\frac{1}{2}\int_{\mathcal V}\left\{\frac{1}{\mu_0}\left[-s^2 \Phi \Delta_{5}\Phi+s^2 {\mathcal T}^2\right]+{%\overline
\rho}\left[V+2{\mathcal U}\right]\right\}{\rm d}{\mathcal V},\nonumber\\
\end{eqnarray}
where ${\mathcal U}$ is the specific internal energy per unit mass \citep{Woltjer:1958, Woltjer:1959b, Broderick:2008}. To obtain the minimal energy equilibrium state for the given invariants ${\mathcal I}_{{\rm I};n}$ and a given helicity and azimuthal flux, we thus minimize $E$ with respect to ${\mathcal I}_{{\rm I};n}$, ${\mathcal I}_{\rm I\!I;0}$ and ${\mathcal I}_{\rm I\!I;1}$. This leads, introducing the associated Lagrangian multipliers $\left(a_{{\rm I};n},a_{{\rm I\!I};0},a_{{\rm I\!I};1}\right)$, to the following condition for a stationary energy:
\begin{equation}
\delta E+\sum_{n}a_{{\rm I};n}\:\delta{\mathcal I}_{{\rm I};n}+\sum_{q=0}^{1}a_{{\rm I\!I};q}\:\delta{\mathcal I}_{{\rm I\!I};q}=0.
\label{Varia1}
\end{equation}
Following the method described in \cite{Chandrasekhar:1958} and \cite{Woltjer:1959b}, %described in Chandrasekhar (1958) and applied in the compressible case by Woltjer (1959b),
we express $\delta E$ and $\delta I_{J;r}$ in function of $\delta\Phi$, $\delta{\mathcal T}$ and $\delta\rho$. Since these variations are independent and arbitrary, their coefficients in the integrand of Eq. (\ref{Varia1}) must separately vanish, which gives
\begin{eqnarray}
\frac{1}{\mu_0}\Delta_{5}\Phi&=&{\overline\rho}\sum_{n}a_{{\rm I};n}\frac{{\rm d}M_{n}\left(s^2 \Phi\right)}{{\rm d}\left(s^2\Phi\right)}+\sum_{q=0}^{1}a_{{\rm I\!I};q}\,{\mathcal T}\,\frac{{\rm d}N_{q}\left(s^2\Phi\right)}{{\rm d}\left(s^2\Phi\right)}\nonumber\\
&=&{\overline\rho}\sum_{n}a_{{\rm I};n}\frac{{\rm d}M_{n}\left(s^2 \Phi\right)}{{\rm d}\left(s^2\Phi\right)}+a_{{\rm I\!I};1}\,{\mathcal T},
\label{RS1}
\end{eqnarray}
\vskip -10pt
\begin{equation}
\frac{1}{\mu_0}s^2\,{\mathcal T}=-\sum_{q=0}^{1}a_{{\rm I\!I};q}\:N_{q}\left(s^2 \Phi\right)=-a_{{\rm I\!I};0}-a_{{\rm I\!I};1} s^2 \Phi.
\label{RS2}
\end{equation}
{These equations thus describe the minimal non force-free energy equilibrium states for a given helicity and azimuthal flux. From Eq. (\ref{GS}), we identify that $F\left(\Psi\right)$ is now constrained while $G\left(\Psi\right)$ {required to ensure the non force-free character of the field} is still arbitrary.}
 
Let us now consider the first invariants family (${\mathcal I}_{\rm I}$) given in Eq. (\ref{I1}) {which are thus necessary to constrain $G\left(\Psi\right)$}. First, the non-magnetic global quantity, which is an invariant of the considered equilibrium, is the total mass of the stellar radiation zone $M_{\rm RZ}$. We thus set $M_{0}\left(s^2 \Phi\right)=1$, leading naturally to consider the mass
\begin{equation}
{\mathcal I}_{{\rm I};0}=\int_{\mathcal V}{\overline\rho}\,{\rm d}{\mathcal V}=M_{\rm RZ}.
\end{equation}
{However, since ${{\rm d}M_{0}\left(s^2 \Phi\right)}/{{\rm d}\left(s^2\Phi\right)}=0$, we have thus to consider the highest-order invariant
\begin{equation}
{\mathcal I}_{{\rm I};1}=\int_{\mathcal V}\left(s^2 \Phi\right){\overline\rho}\,{\rm d}{\mathcal V}
\end{equation}
because of the non force-free behaviour of the field. This last invariant has been introduced by \cite{Prendergast:1956} in the axisymmetric non force-free magneto-hydrostatic incompressible equilibrium and it corresponds to the mass conservation in each flux tube described by the closed magnetic surface $s^2 \Phi={\rm c}^{\rm ste}$.

Furthermore, the considered radiation zone is stably stratified. Since in stellar interiors the magnetic pressure is very much less than the thermal one, the Lorentz force have only a negligible effect on the gas pressure ($\beta\!>\!\!>\!1$). Moreover, energy is required to move fluid elements in the radial direction because work has to be done against the buoyant restoring force which is thus very strong compared to the magnetic one. Therefore, the radial component of the displacement ($\vec \xi$) which takes place during the adjustment to equilibrium is inhibited $\vec\xi\cdot{\widehat{\bf e}}_{r}\approx0$ and $\vec\nabla\cdot\left({\overline\rho}{\vec\xi}\right)\approx0$ due to the anelastic approximation justified in stellar radiation regions. Therefore, as emphasized by \cite{Braithwaite:2008M}, the mass transport in the radial direction is frozen (no matter can leave or enter in the flux tube) and ${\mathcal I}_{{\rm I};1}$ can be used as a supplementary constrain in our variational method.}\\

From Eqs. (\ref{RS1}-\ref{RS2}), we thus get the following equations describing the barotropic axisymmetric equilibrium state of lowest energy that the compressible star can reach while conserving its radiation zone mass (${\mathcal I}_{{\rm I};0}=M_{\rm RZ}$), the mass in each flux tube (${\mathcal I}_{{\rm I};1}$), the flux of the toroidal field (${\mathcal I}_{{\rm I\!I};0}={\mathcal F}_{\varphi}$), and a given helicity (${\mathcal I}_{{\rm I\!I};1}={\mathcal H}/2$):
\begin{equation}
\frac{1}{\mu_0}\Delta_{5}\Phi=a_{\rm I;1}\:{\overline \rho}+a_{\rm I\!I;1}\:{\mathcal T},
\label{RS3}
\end{equation}
\begin{equation}
\frac{1}{\mu_0}{\mathcal T}=-a_{\rm I\!I;1}\Phi-\frac{a_{\rm I\!I;0}}{s^2}.
\label{RS4}
\end{equation}
Since the azimuthal field is regular at the origin, we get from Eq. (\ref{Cy2}) $a_{{\rm I\!I};0}=0$. Eliminating ${\mathcal T}$ between Eqs. (\ref{RS3}-\ref{RS4}), we obtain 
\begin{equation}
\Delta_{5}\Phi+ \left[\mu_{0}a_{\rm I\!I;1}\right]^2\Phi=\mu_{0}\:a_{\rm I;1}\:{\overline \rho}
\end{equation}
that becomes, multiplying it by $s^2$ and using Eq. (\ref{Phitau1}) \& (\ref{Phitau2})
\begin{equation}
\Delta^{*}\Psi+\left[\mu_{0}a_{\rm I\!I;1}\right]^2\Psi=\mu_{0}\:a_{\rm I;1}\:{\overline\rho}\:r^2\sin^2\theta.
\end{equation}
We thus identify in Eq. (\ref{GSNL})
\begin{equation}
\left \{
\begin{array}{c @{=} c}
    k & 1 \\
    j & 0 \\
\end{array}
\right.\quad\hbox{and}\quad
\left \{
\begin{array}{c @{=} c}
    a_{\rm I;1} & - \beta_{0} \\
    a_{{\rm I\!I};1} & -\frac{1}{\mu_0}{\lambda_1}/R\\
\end{array}
\right.,
\end{equation}
where we have constrained the initial arbitrary functions of the magnetohydrostatic equilibrium
\begin{equation}
F\left(\Psi\right)=-\mu_{0}a_{{\rm I\!I};1}\Psi\quad\hbox{and}\quad G\left(\Psi\right)=-%\mu_{0}
a_{\rm I;1}.
\end{equation}}
So, it reduces to
\begin{equation}
\Delta^{*}\Psi +\frac{\lambda_1^2}{R^2}\, \Psi = - \mu_0\,\overline\rho\, r^2 \sin^2 \theta\,\beta_0,
\label{GSDM}
\end{equation}
the values of the real coefficients $\lambda_1$ and $\beta_0$ being {thus controled by the helicity (${\mathcal H}$) and the mass conservation in each axisymmetric flux tube defined by $\Psi={\rm c}^{\rm ste}$ {because of the non force-free stably stratified behaviour of the reached equilibrium}. %It thus appears since ${\mathcal I}_{\rm I;1}\ne0$ in the self-gravitating case that $\beta_0\ne0$ and thus that the field is not force-free in the general self-gravitating axisymmetric case.}

{This corresponds, as has already been emphasized, to the lowest energy equilibrium state for a given helicity 
\citep{Bellan:2000,Broderick:2008}.
%(Bellan 2000 and Broderick \& Narayan 2008).
The equilibrium state ruled by Eq. (\ref{GSDM}) are thus the generalization of the Taylor relaxation states in a self-gravitating star where the field is not force-free ({\it i.e.} $\bnab \times\mbf B\ne\alpha\:\mbf B$). {Note also that some non force-free relaxed states have been identified in plasma physics \citep{MontgomeryPhillips:1989A, MontgomeryPhillips:1989D, Dasguptaetal:2003, Shaikhetal:2008} that  should be studied in a stellar context in a near future.}

Let us note that in the case where ${\mathcal I}_{\rm I;1}$ is not considered ($a_{\rm I;1}=\beta_0=0$) we recover the \cite{Chandrasekhar:1956a} {force-free} limit \citep[see also][for a generalization of the solutions]{Marsh:1992} and the usual Taylor's states for low-$\beta$ plasmas. {The Prendergast's model is recovered assuming a constant density profile (incompressible).}\\

\subsection{Green's function solution}
\noindent We are now ready to solve Eq. (\ref{GSDM}). If we introduce $x=\cos\theta$ and if we set ${\mathcal S}\left(r,\theta\right)=-\mu_0\:\beta_0 \:\overline\rho\: r^2\sin^2\theta$, it is recast in
\begin{equation}
{\mathcal L}_{\lambda_1}\Psi={\mathcal S},
\label{inhom}
\end{equation}
where
\begin{equation}
{\mathcal L}_{\lambda_1}\equiv\left[\partial_{rr}+\frac{1-x^2}{r^2}\partial_{xx}+\frac{\lambda_{1}^{2}}{R^2}\right].
\end{equation}
Using Green's function method 
\citep{Morse:1953}
%(Morse \& Feshbach 1953)
, we then obtain {the particular solution associated with $\mathcal S$}:
\begin{eqnarray}
\lefteqn{\Psi_{\rm p}\left(r,x\right)=-\mu_0\,\beta_0\sum_{l}\frac{\lambda_{1}^{l}}{R_{\rm sup}}\left[\frac{2l+3}{2\left(l+1\right)\left(l+2\right)}\right]\times}\nonumber\\
&&\left\{r\, j_{l+1}\left(\lambda_{1}^{l}\,\frac{r}{R_{\rm sup}}\right)\int_{r}^{R_{\rm sup}}\left[
\xi\,y_{l+1}\left(\lambda_{1}^{l}\,\frac{\xi}{R_{\rm sup}}\right){\mathcal J}_{l}\left(\xi\right)\right]{\rm d}\xi\right.\nonumber\\
&&{\left. +r\, y_{l+1}\left(\lambda_{1}^{l}\,\frac{r}{R_{\rm sup}}\right)\int_{R_{\rm inf}}^{r}\left[
\xi\,j_{l+1}\left(\lambda_{1}^{l}\,\frac{\xi}{R_{\rm sup}}\right){\mathcal J}_{l}\left(\xi\right)\right]{\rm d}\xi\right\}}\nonumber\\
&&\times\left(1-x^2\right)C_{l}^{3/2}\left(x\right),
\end{eqnarray}
where
\begin{equation}
{\mathcal J}_{l}\left(\xi\right)=\int_{-1}^{1}{\mathcal S}\left(\xi,x'\right)C_{l}^{3/2}\left(x'\right){\rm d}x';
\end{equation}
$j_{l}$ and $y_{l}$ are respectively the spherical Bessel functions of the first and the second kind (also called Neumann functions) while $C_{l}^{3/2}$ are the Gegenbauer polynomials 
\citep{Abramowitz:1972}. 
%(see Abramowitz \& Stegun 1972). 
{$R_{\rm inf}$ and $R_{\rm sup}$, which are respectively the bottom and the top radius of the considered radiation zone, are introduced and we identify $R=R_{\rm sup}$.}

These functions ($j_{l}$, $y_{l}$, and $C_{l}^{3/2}$) are respectively the radial and the latitudinal eigenfunctions of the homogeneous equation associated with Eq. (\ref{inhom}) :
\begin{equation}
{\mathcal L}_{\lambda_1}\Psi_{\rm h}=0.
\label{hom}
\end{equation}
Then, if we express the solutions of this equation as $\Psi_{\rm h}=\sum_{l}f_{l}\left(r\right)g_{l}\left(\theta\right)$, we get respectively
\begin{equation}
\left(1-x^2\right)\frac{{\rm d}^2 g_{l}}{{\rm d}x^2}+\left(l+1\right)\left(l+2\right)g_{l}=0
\end{equation}
and
\begin{equation}
\frac{{\rm d}^2 f_l}{{\rm d}r^2}+\left[\left(\frac{\lambda_{1}^{l}}{R_{\rm sup}}\right)^2-\frac{\left(l+1\right)\left(l+2\right)}{r^2}\right]f_{l}=0\,,
\end{equation}
giving 
\begin{equation}
g_{l}=\left(1-x^2\right)C_{l}^{3/2}\left(x\right)
\end{equation}
 and 
\begin{eqnarray}
f_{l}&=&K_{1}^{l}\,\lambda_{1}^l\frac{r}{R_{\rm sup}}\,j_{l+1}\left(\lambda_{1}^l\frac{r}{R_{\rm sup}}\right)\nonumber \\
&&+K_{2}^{l}\,\lambda_{1}^l \frac{r}{R_{\rm sup}}\,y_{l+1}\left(\lambda_{1}^l\frac{r}{R_{\rm sup}}\right),
\end{eqnarray}
$K_{1}^{l}$ and $K_{2}^{l}$ being real constants. $\lambda_{1}^{l}$ are eigenvalues that allow to verify the boundary conditions discussed hereafter.
One has to notice that $K_{2}^{l}$ has to vanish in order to preserve the regularity of the solution at the center.
Applying this to Eq. (\ref{GSDM}), we finally obtain the general solution
\begin{eqnarray}
&&\Psi\left(r,\theta \right)=\Psi_{\rm h}+\Psi_{\rm p}\nonumber\\
&=&
\sin^2\theta\,\times \Bigg\{ \sum_{l=0}^{\infty}K_1^l\frac{\lambda_{1}^{l,i}}{R_{\rm sup}}\,r\, j_{l+1} \left(\lambda_1^{l,i}\frac{\: r}{R_{\rm sup}}\right) C_{l}^{3/2}\left(\cos \theta\right)
\nonumber\\
&-&
\mu_0\beta_0\frac{\lambda_{1}^{0,i}}{R_{\rm sup}}r j_{1}\left(\lambda_{1}^{0,i}\frac{r}{R_{\rm sup}}\right)\int_{r}^{R_{\rm sup}}\!\left[y_{1}\left(\lambda_{1}^{0,i}\,\frac{\xi}{R_{\rm sup}}\right)\overline\rho \xi^3\right]\!{\rm d}\xi
\nonumber\\
&-&
\mu_0\beta_0\frac{\lambda_{1}^{0,i}}{R_{\rm sup}}r y_{1}\left(\lambda_{1}^{0,i} \frac{r}{R_{\rm sup}}\right)\!\int_{R_{\rm inf}}^{r}\left[
j_{1}\left(\lambda_{1}^{0,i}\frac{\xi}{R_{\rm sup}}\right)\overline\rho\xi^3\!\right]\!{\rm d}\xi
\Bigg\}.\nonumber\\
\end{eqnarray}
%Note that $\Psi_{\rm p}$ presents a dipolar geometry due to angular dependance of the source term $\mathcal S=-\mu_0\:\beta_0\:\overline\rho\: r^2\sin^2\theta$. 
%
%The field is thus given for $r\le R_{\rm sup}$ by:
%\begin{equation}
%\mbf B=\frac{1}{r^2\sin\theta}\partial_{\theta}\Psi\,\er-\frac{1}{r\sin\theta}\partial_{r}\Psi\,\etheta+\frac{\lambda_{1}^{0,i}}{R_{\rm sup}}\frac{\Psi}{r\sin\theta}\,\ephi.
%\label{BPsi}
%\end{equation}
{Notice that the particular solution for the poloidal flux function $\left(\Psi_{\rm p}\right)$ presents a dipolar geometry, owing to its angular dependence that follows the one from the source term $\mathcal S=-\mu_0\:\beta_0\:\overline\rho\: r^2\sin^2\theta$. \\ 
Neglecting the density, we end up with the linear homogeneous equation, whose solutions are Chandrasekhar-Kendall functions \citep{Chandrasekhar:1957}. Moreover, the source term being constituted only of a dipolar component, all the non-dipolar contributions are, according to this model, force-free.\\ 

\noindent The magnetic field is then given for $r\le R_{\rm sup}$ by
\begin{equation}
\mbf B=\frac{1}{r^2\sin\theta}\partial_{\theta}\Psi\,\er-\frac{1}{r\sin\theta}\partial_{r}\Psi\,\etheta+\frac{\lambda_{1}^{0,i}}{R_{\rm sup}}\frac{\Psi}{r\sin\theta}\,\ephi.
\label{BPsi}
\end{equation}
After a few manipulations, we can then express the current density as
\begin{eqnarray}
\mbf{j}_{\rm P} &=\frac{1}{\mu_0} \bnab \times \mbf{B}_{\rm T} &= \underbrace{\frac{\lambda_1^{0,i}}{\mu_0 \: R} \: \mbf{B}_{\rm P}}_{\rm force-free},
\label{jPalphaBT}\\
\mbf{j}_{\rm T} &=\frac{1}{\mu_0} \bnab \times \mbf{B}_{\rm P} &= \underbrace{\frac{\lambda_1^{0,i}}{\mu_0 \: R} \:  \mbf{B}_{\rm T}} _{\rm force-free}+\underbrace{ \beta_0 \:Ê{\overline\rho} \: r \sin\theta \: \ephi}_{\rm non\:force-free},
\label{jTalphaBP}
\end{eqnarray}
where we recognize in the first term of the right hand side the force-free contributions and in the second the non force-free one, fully contained in the toroidal component.

The Lorentz force can as a matter of fact be written {\bf in} the very simple form: 
\begin{equation}
\mbf{F}_{\mathcal{L}} = \mbf{F}_{\mathcal{L}_{\rm P}} = \beta_0 \:Ê{\overline\rho} \: \bnab \Psi .
\label{FLPsi}
\end{equation}}
\subsection{Configurations}

The boundary conditions for $\Psi$ which {determine} possible values for $K_1^l$ and $\lambda_{1}^{0,i}$ have now to be discussed. 
Two major types of geometry are relevant for large-scale fossil magnetic fields in stellar interiors: initially confined and open configurations.\\

\subsubsection{Initially confined configurations}

Let us first concentrate on the simplest mathematical solution in the case of a central radiation zone that initially cancels $\Psi$ both at the center ($R_{\rm inf}=0$) and at a given confinement radius ($R_{\rm sup}=R_c$). 

Then, if we choose to cancel the $K_{1}^l$ coefficients for every $l$, the condition $\Psi(0,\theta)=0$ is verified since $\lim_{r\to0}r\, j_{1}\left(\lambda_{1}^{0,i}\frac{r}{R_{\rm sup}}\right)=0$.
%Next, two choices for the upper boundary condition where $\Psi\left(R_c,0\right)=0$ (i.e. $\mbf B\cdot{\widehat{\mbf e}}_r=0$) can be adopted. The first one is such that $\Psi\left(R_c,\theta\right)$ vanishes with $B_{r}\left(R_c,\theta\right)=0$ and $B_{\theta}\left(R_c,\theta\right)\ne0$. Then, $\lambda_{1}^{0,n}=\alpha_{1,n}$, where $\alpha_{1,n}$ are the roots of the Neumann function $y_1$. The second most stringent choice is such that $\mbf B\left(R_c,\theta\right)=\mbf 0$; then, $\partial_{\theta}\Psi\left(R_c,\theta\right)=\partial_{r}\Psi\left(R_c,\theta\right)=0$, giving $\int_{0}^{R_c}\!\left[\,j_{1}\!\!\left(\lambda_{1}^{0,i}\,\frac{\xi}{R_c}\right)\!\overline\rho\,\xi^3\right]\!{\rm d}\xi=0$. 
However, if we look at the magnetic field radial component behaviour at the center, it is easily shown, using Eq. (\ref{BPsi}), that if $K_{1}^{0}=0$ it is given by $\lim_{r\to0}r^{-1}j_{1}\left(\lambda_{1}^{0,i}\frac{r}{R_{\rm sup}}\right)$ which does not cancel so that $B_{r}\left(0,\theta\right)\ne0=C\,\cos\theta$ (where $C\in\mathbb{R}^{*}$). Therefore, this solution is multivaluated, thus physically inadmissible and $K_{1}^{0}\ne0$.\\

Then, we consider the general case of a field initially confined between two radii $R_{\rm inf}=R_{c_1}$ and $R_{\rm sup}=R_{c_2}$, owing to the presence of both a convective core and a convective envelope (as it is the case e.g. in A-type stars). We impose $\Psi\left(R_{c_1},\theta\right)=0$ and $\Psi\left(R=R_{c_2},\theta\right) = 0$, which gives the two independent equations for $l=0$
\begin{equation}
%\lefteqn{
K_1^0%\,j_1\left(\lambda_1^{0,i}\: \frac{R_{c_1}}{R_{c_2}}\right)}\nonumber\\
=
\mu_0\,\beta_0
%j_{1}\!\!\left(\lambda_{1}^{0,i}\,\frac{R_{c_1}}{R_{c_2}}\right)\!
\int_{R_{c_1}}^{R_{c_2}}\!\left[
\,y_{1}\left(\lambda_{1}^{0,i}\,\frac{\xi}{R_{c_2}}\right)\overline\rho\,\xi^3\right]{\rm d}\xi\,\,
\end{equation}
and
\begin{eqnarray}
K_1^0 j_1\left(\lambda_1^{0,i}\right) \!=\!  \mu_0\,\beta_0y_1\!\left(\lambda_1^{0,i}\right)\!
\int_{R_{c_1}}^{R_{c_2}}\!\left[\,
j_{1}\left(\lambda_{1}^{0,i}\frac{\xi}{R_{c_2}}\right)\overline\rho\xi^3\right]{\rm d}\xi;\;\;
\label{K0}
\end{eqnarray}
we here focus on the dipolar mode which is known to be the lowest energy per helicity ratio state \citep[cf.][]{Broderick:2008}.
These can be formulated so that one first determines the value of $\lambda_{1}^{0,i}$ according to
\begin{eqnarray}
&& j_1\left(\lambda_1^{0,i}\right) \: 
\int_{R_{c_1}}^{R_{c_2}}\!\left[
\,y_{1}\left(\lambda_{1}^{0,i}\,\frac{\xi}{R_{c_2}}\right)\overline\rho\,\xi^3\right]{\rm d}\xi \nonumber\\
&&-y_1\left(\lambda_1^{0,i}\right)\: 
\int_{R_{c_1}}^{R_{c_2}}\!\left[
\,j_{1}\left(\lambda_{1}^{0,i}\,\frac{\xi}{R_{c_2}}\right)\overline\rho\,\xi^3\right]{\rm d}\xi=0\label{L1}\end{eqnarray}
and next computes $K_1^0$ following (\ref{K0}).

In the case where there is no convective core, as for example in central radiation zones of late-type stars such as the Sun, Eqs. (\ref{K0}) \& (\ref{L1}) must be applied setting $R_{c_1}=0$.\\

\subsubsection{Open configurations}

This corresponds to the case of fields that match at the stellar surface (at $r=R_{*}$, $R_{*}$ being the star's radius) with a potential field as observed now in some cases of early-type stars such as Ap stars. Then, we have ${\mbf B}_{\rm ext}=\bnab \Phi_{\rm M}$, $\Phi_{\rm M}$ being the associated potential.\\
%$\lambda_{1}^{0,k}=k\pi/2\,(k\in \mathbb{N}^{*})$.
%Let us emphasize that in this case $B_{\varphi}=0$ for $r>R_{*}$ due to the strong Ohmic diffusivity, which cancels $\mbf j$ {and that} since $F=\lambda_{1}^{0,i}\Psi$ for $r\le R_{*}$, we get a discontinuity for $B_{\varphi}$ at the surface, which is however allowed by the Maxwell-Amp\`ere's equation.

In the case studied here, we focus on the first configuration (initially confined) since the search of relaxed solutions for given ${\mathcal I}_{{\rm I};0}$, ${\mathcal I}_{{\rm I};1}$, ${\mathcal I}_{{\rm I \! I};0}$, and ${\mathcal I}_{{\rm I\! I};1}$ assumes that $\bB\cdot\er=0$ on the stellar radiation zones boundaries. This initial confined configuration will then become open one 
%leads to the first previous confined configuration, that can be seen as a potential prelude to observed multipolar one, this latter stage being achieved 
through Ohmic diffusion as in the Braithwaite and collaborators scenario.
%
%****************************************************************************
%***************     Section : Application to Realistic Stellar Interiors ****** 
%****************************************************************************
\section{Application to Realistic Stellar Interiors}
To illustrate our purpose, we apply our analytical results (i) to model an initial fossil field buried below the convective envelope of the young Sun on the ZAMS, and then (ii) to model an initial field present in the radiation zone of a ZAMS $2.40\: M_\odot$ magnetic Ap-star, whose lower and upper radiation-convection interfaces are respectively located at $R_{c_1}=0.111 \:R_*$ and at $R_{c_2}=0.992\: R_*$. In the first case, the parameter $\beta_0$ is determined such that the maximum field strength reaches the amplitude of $B_0=2.1\, {\rm MG}$, which is one of the upper limits given by \cite{Friedland:2004} for the present Sun's radiative core. In the second case, it is obtained such that it reaches the arbitrary value of $B_0=10\, {\rm kG}$. This value is approximatively of the same order of magnitude that the mean surface amplitude observed using spectropolarimetry for magnetic Ap-star which exhibits strong external dipolar magnetic behaviour (such as HD12288 \citep{Wade:2000}). We thus assume implicitly that such an initial confined internal field is a potential prelude to the multipolar one observed now at the surface, the latter state being acheived after a diffusive process that will be studied in a forthcoming paper.
\subsection{Fossil fields buried in late-type stars radiative cores}
The young Sun model used as a reference is a \textsc{Cesam} non-rotating standard one 
\citep{Morel:1997}, 
following inputs from the work of \cite{Couvidat:2003} and \cite{Turck-Chieze:2004}.

In Fig. \ref{Sol1}, three possible configurations for $\Psi$ are given. We choose those corresponding to the first, the third and the fifth eigenvalues (given in table \ref{eigenvals}). Those are the generalization of the well-known Grad-Shafranov equation linear eigenmodes obtained in the force-free case \citep[cf.][]{Marsh:1992}. The field is then of mixed-type ($B_{\varphi}$ is given for $\lambda_{1}^{0,1}$), both poloidal and toroidal, and non force-free, properties already obtained by \cite{Prendergast:1956} in the incompressible case. The respective amplitudes ratio between the poloidal and the toroidal components will be described in \S 5., where the {possible} stability of such configurations will be discussed.

%--- Table1 ---
\begin{table}[h!]
\caption{Eigenvalues of the first five equilibria for the two configurations illustrated.} % title of Table
\label{eigenvals} % is used to refer this table in the text
\centering % used for centering table
\begin{tabular}{c c c } % centered columns (3 columns)
\hline\hline % inserts double horizontal lines
Eigenvalue & Solar case & Ap star case \\
\hline
$\lambda_1^{0,1}$ & 5.276 & 4.826\\
$\lambda_1^{0,2}$ & 9.157 & 8.657\\
$\lambda_1^{0,3}$ & 12.951 & 12.444\\
$\lambda_1^{0,4}$ & 16.290 & 16.174\\
$\lambda_1^{0,5}$ & 19.839 & 19.849\\
\hline %inserts single line
\end{tabular}
\end{table}

\begin{figure*}[tp!]
\begin{center}
\includegraphics[width=0.35\textwidth]{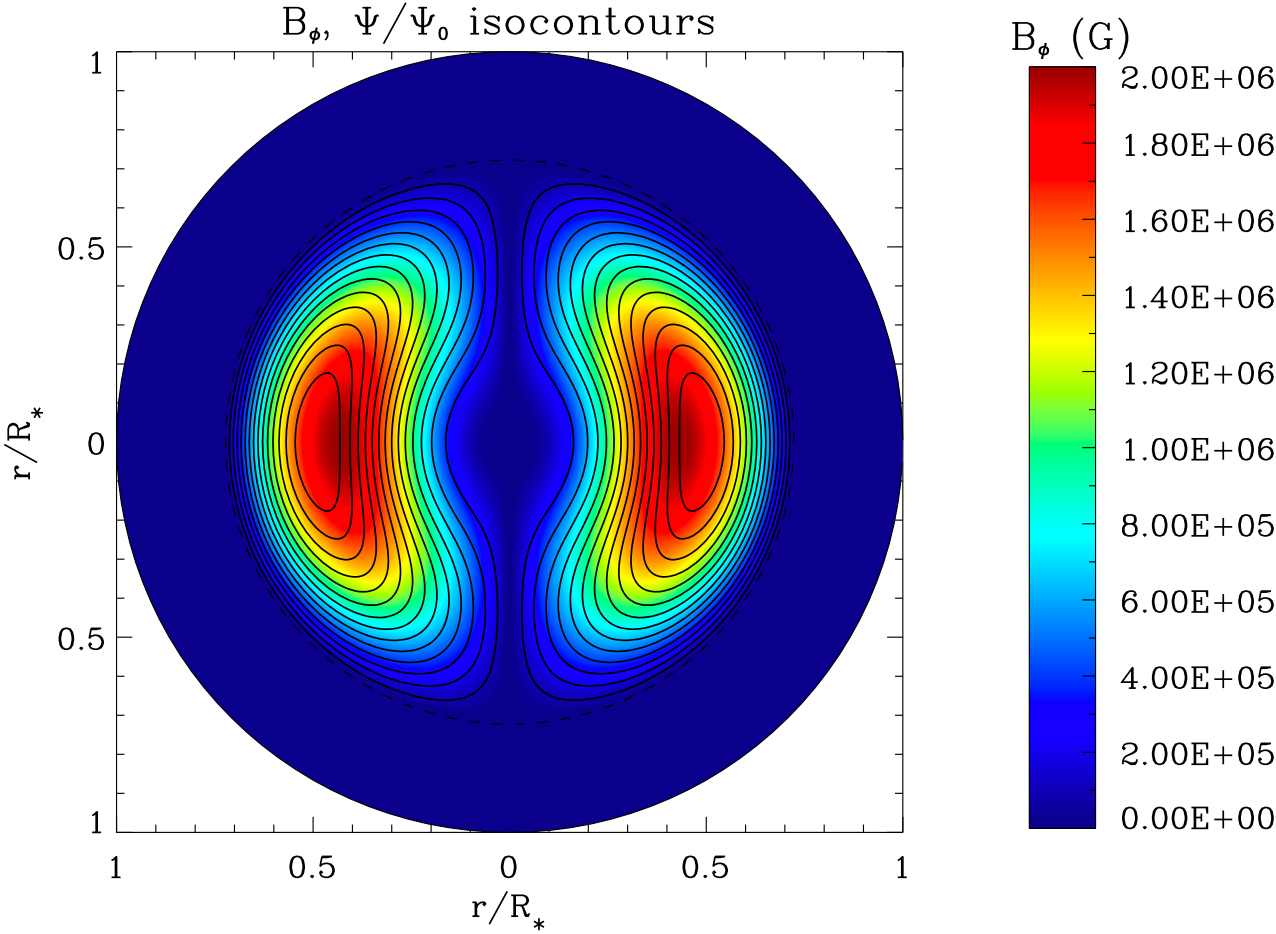}
\includegraphics[width=0.35\textwidth]{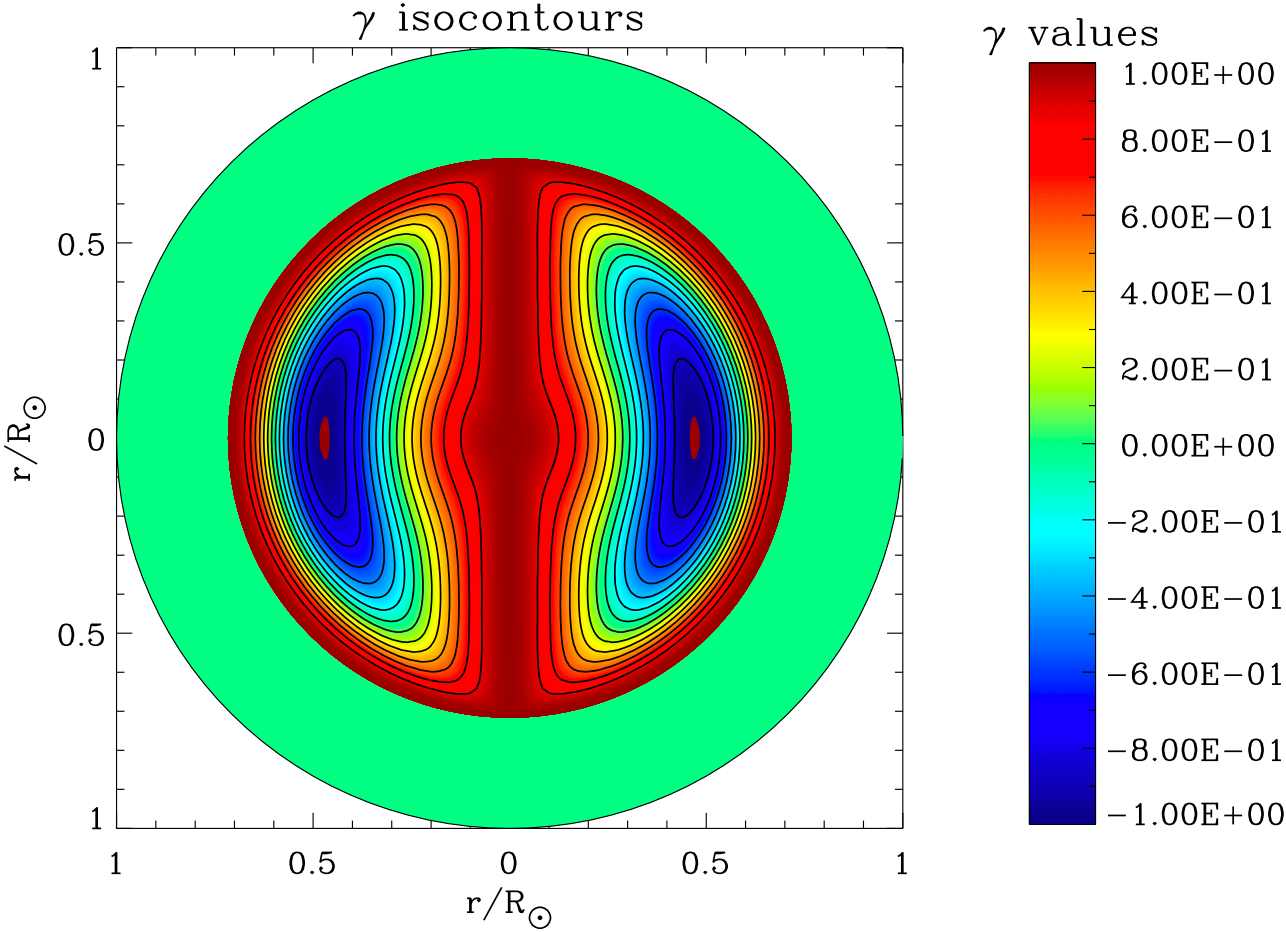}\\
\includegraphics[width=0.35\textwidth]{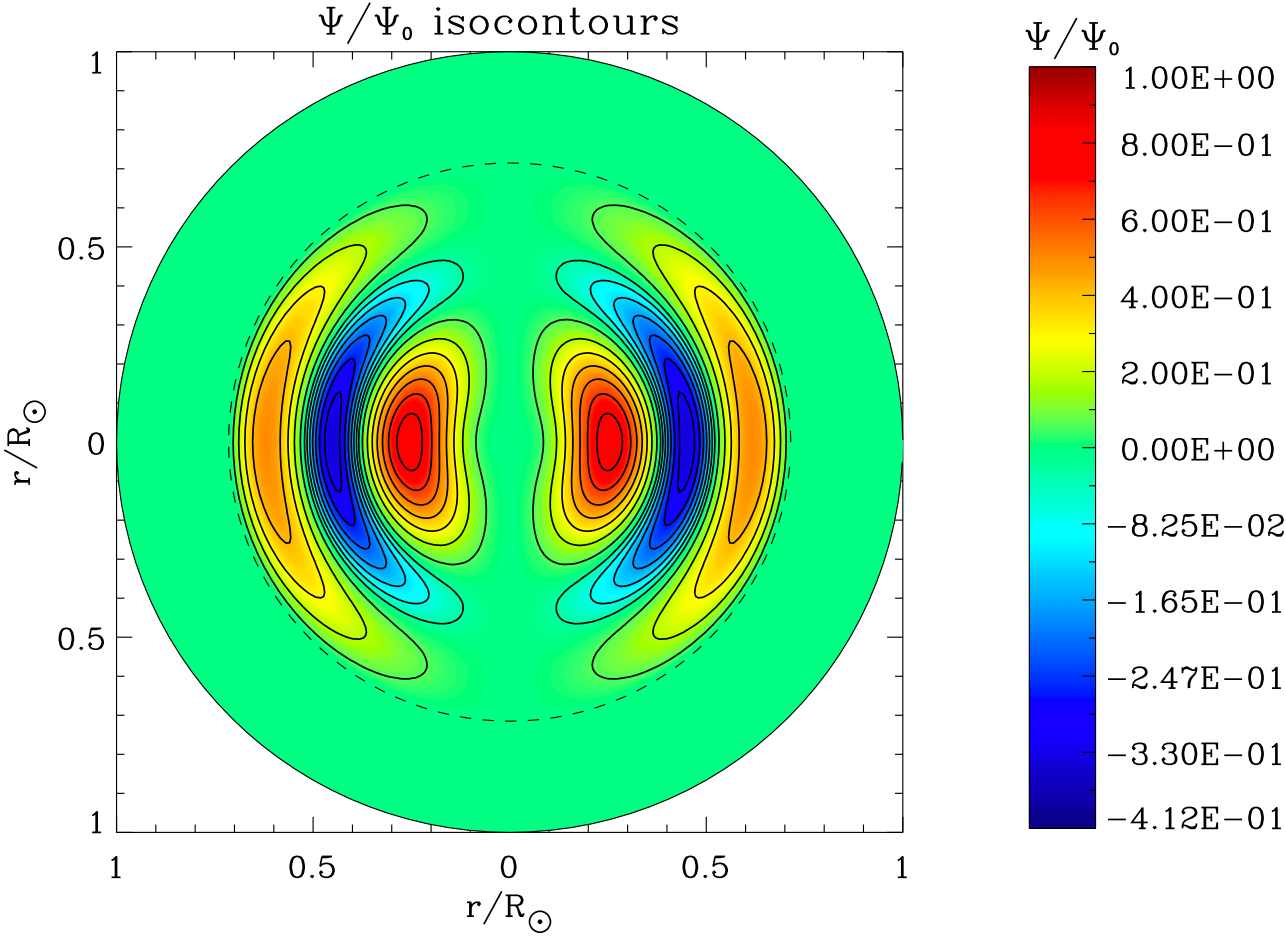}
\includegraphics[width=0.35\textwidth]{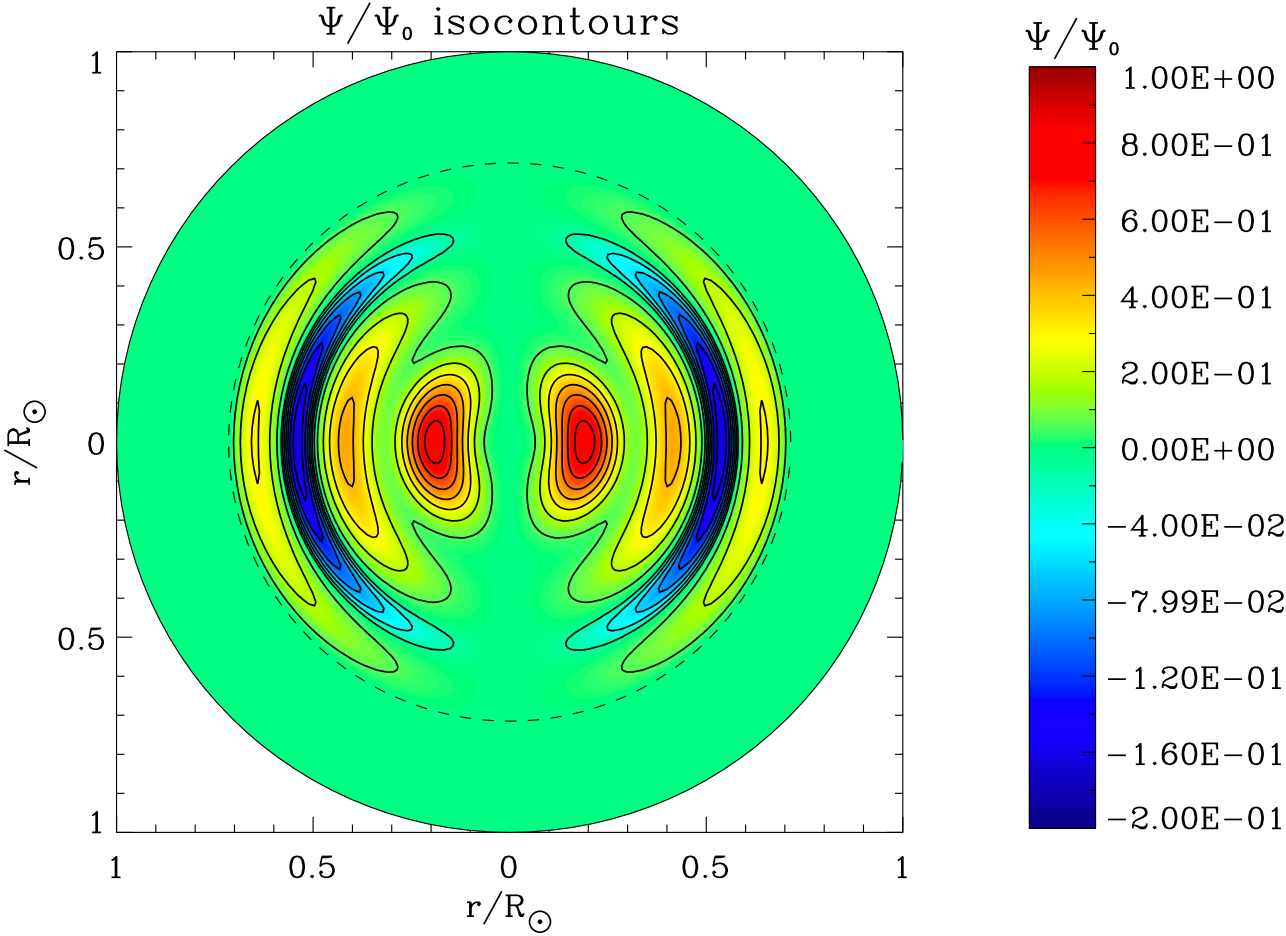}
\end{center}
\vskip -15pt
\caption{Upper panel: (left) toroidal magnetic field strength in colorscale and normalized isocontours of the poloidal flux function $\Psi$ in meridional cut in the solar case where the field is buried in the radiative core (below $0.726 R_*$) for the first equilibrium configuration ($\lambda_{1}^{0,1}$); (right) anisotropy factor $\gamma$ (Eq. \ref{Anis}). Lower panel: normalized isocontours of the flux function $\Psi$ in meridional cut (left) for the third possible eigenvalue ($\lambda_1^{0,3}$) in the same case, (right) for the fifth possible eigenvalue ($\lambda_1^{0,5}$). The dashed circles indicate the radiation-convection limits.
\label{Sol1}}
\end{figure*}
\begin{figure*}[tp!]
\begin{center}
\includegraphics[width=0.35\textwidth]{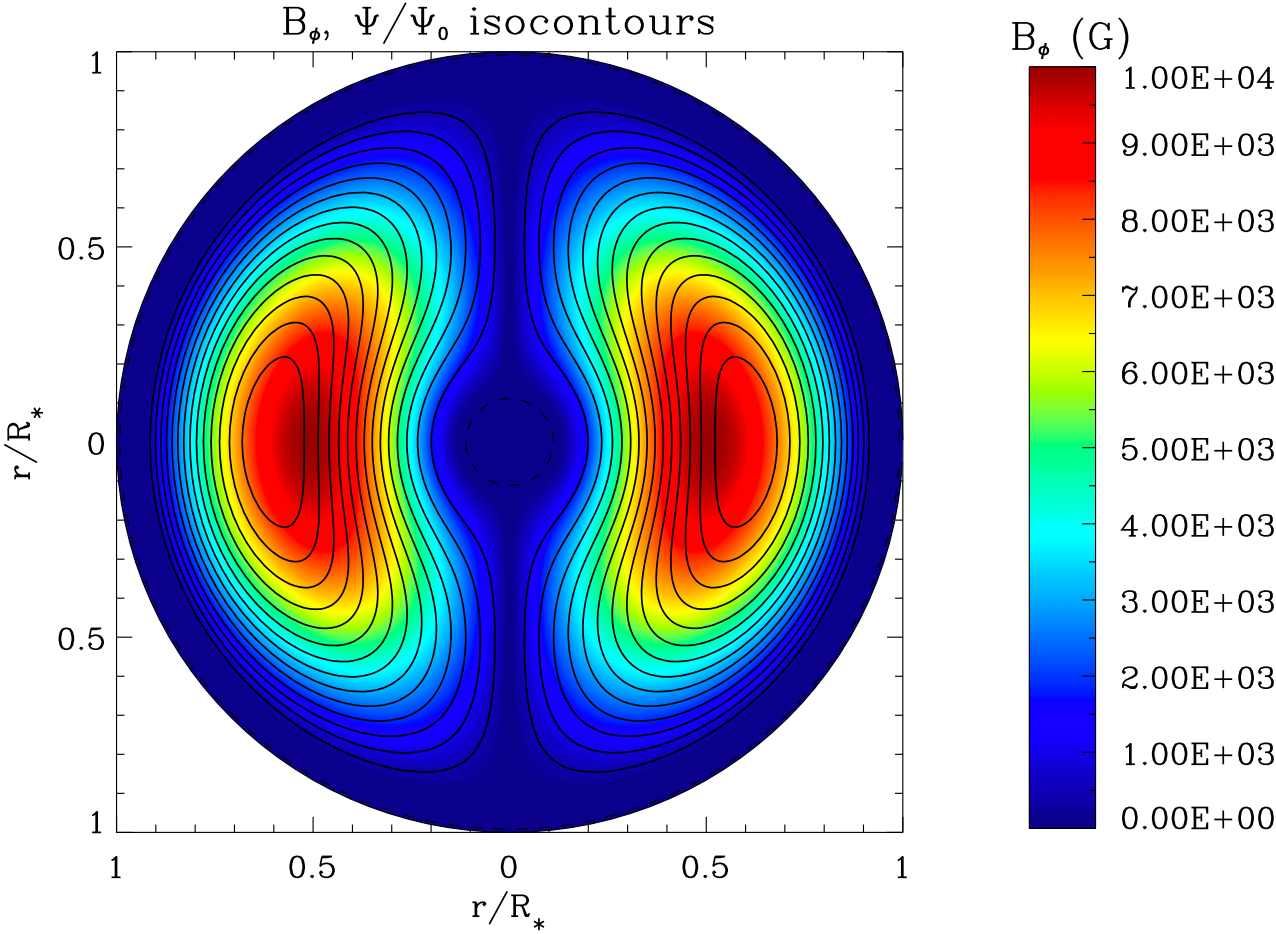}
\includegraphics[width=0.35\textwidth]{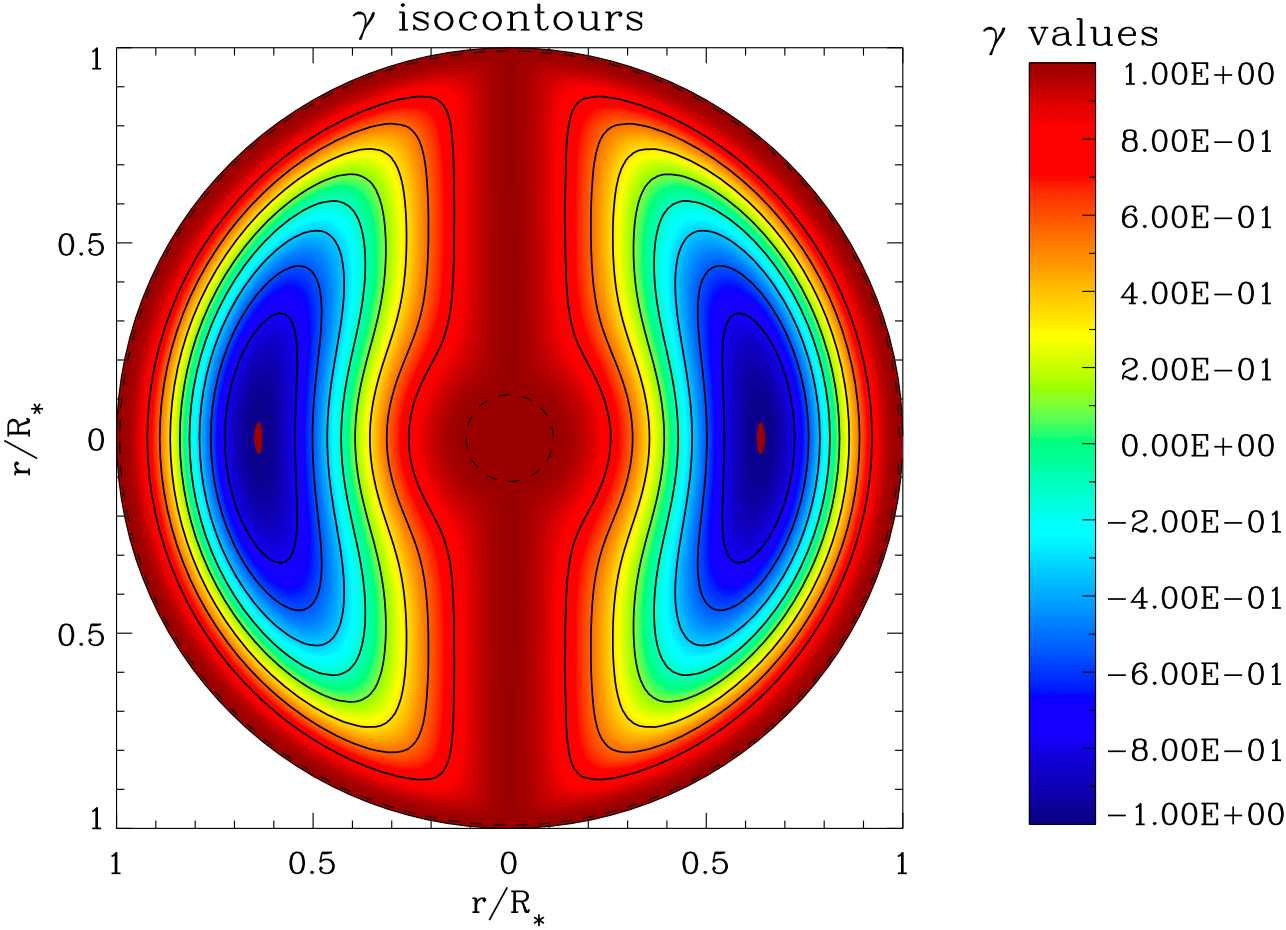}\\
\includegraphics[width=0.35\textwidth]{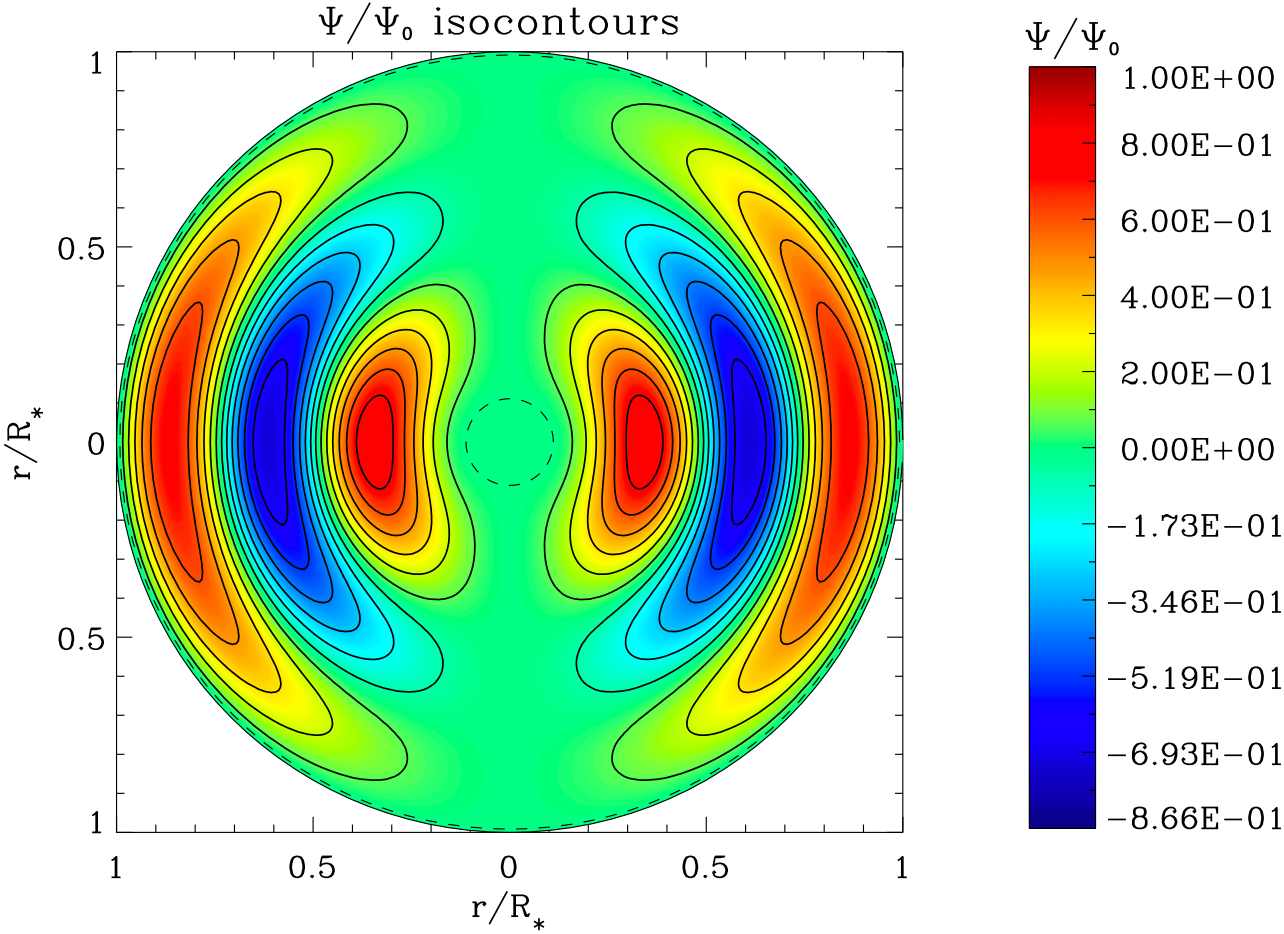}
\includegraphics[width=0.35\textwidth]{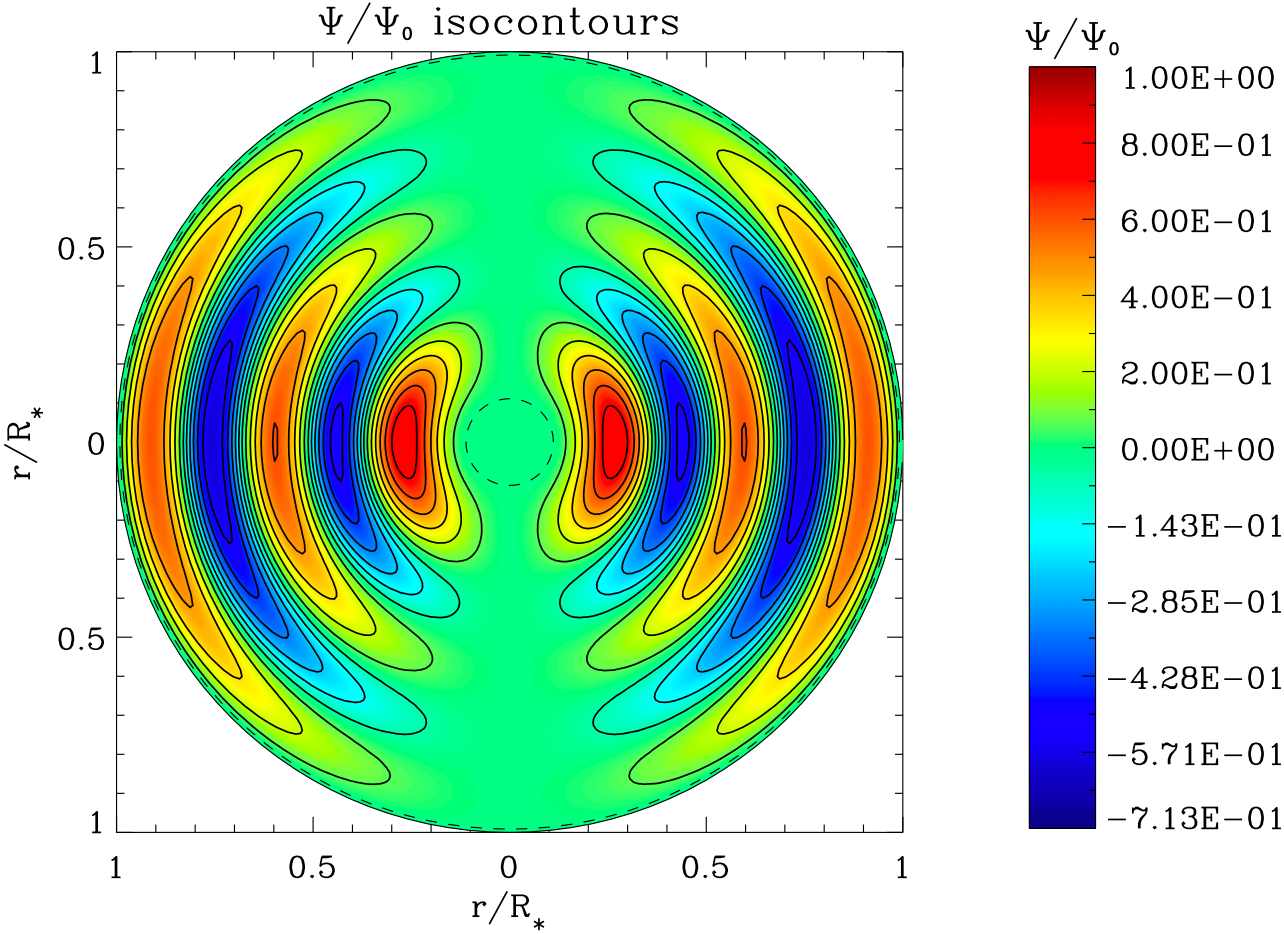}
\end{center}
\vskip -15pt\caption{Upper panel: (left) toroidal magnetic field strength in colorscale and normalized isocontours of the poloidal flux function $\Psi$ in meridional cut in the Ap star's case where the field is confined between $R_{c_1}=0.111\:R_*$ and $R_{c_2}=0.992\:R_*$ for the first equilibrium configuration ($\lambda_{1}^{0,1}$); (right) anisotropy factor $\gamma$ (Eq. \ref{Anis}). Lower panel: normalized isocontours of the flux function $\Psi$ in meridional cut (left) for the third possible eigenvalue ($\lambda_1^{0,3}$) in the same case, (right) for the fifth possible eigenvalue ($\lambda_1^{0,5}$). The dashed circles indicate the radiation-convection limits.
\label{Ap1}}
\end{figure*}
%\newpage
%
\subsection{Fossil fields in early-type stars}
{Respective corresponding possible configurations in the case of an 
%where the field matches with a potential one at the 
Ap star %'s surface
are given in Fig. \ref{Ap1}.} 

The model is typical of an A2p-type star, with an initial mass $M_{\rm A} = 2.40\,\rm{M}_\odot$. The solar metallicity is chosen as the initial one and the model is taken on the ZAMS, its luminosity being $L_*= 38.0 \: \rm{L}_\odot$.\\

Obtained configurations are then mixed poloidal-toroidal (twisted) fields which may be stable in stellar radiation zones  \citep[cf.][]{Braithwaite:2004, Braithwaite:2006}.
Their configurations are thus given in both cases by concentric torus, the neutral poloidal points (where $\partial_{r}\Psi=\partial_{\theta}\Psi=0$ so that $B_r=B_\theta=0$) positions being function of the internal density profile of the star. 

%Note also that the eigenmode that have the lowest eigenvalue ($\lambda_{1}^{0,1}$) is the one that has the minimum $U_{\rm mag}/{\mathcal H}$ ratio (where $U_{\rm mag}=\frac{1}{2\,\mu_{0}}\int_{\mathcal V}{\mbf B}^2{\rm d}{\mathcal V}$) so that it is the preferential mode \citep{Broderick:2008}.\\
Let us emphasize here that the original approach of this work first consists in deriving the Grad-Shafranov-like equation adapted to treat the barotropic magnetohydrostatic equilibrium states for realistic models of stellar interiors. Such approach has already been applied to investigate the internal magnetic configurations in polytropes and in compact objects such  that white dwarfs or neutron stars 
%see for example Monaghan 1976, Payne \& Melatos 2004, Tomimura \& Eriguchi 2005, Yoshida et al. 2006, Haskell et al. 2008, Akg\"un \& Wasserman 2008, Kiuchi \& Kotake 2008). 
\citep[see e.g.][]{Monaghan:1976, Payne:2004, Tomimura:2005, Yoshida:2006, Haskell:2008, Akgun:2008, Kiuchi:2008}.
 
Then, the obtained arbitrary functions are constrained with deriving minimal energy equilibrium configurations for a given conserved mass, azimuthal flux and helicity that generalizes the relaxation Taylor's states to the self-gravitating case where the field is non-force free.

\section{Links between the field's helicity, topology and energy}
\subsection{Helicity {\it vs.} mixity}
\label{heli_mix}
Let us express the magnetic field $\left(\bB\right)$ in terms of magnetic stream functions $\xi_{\rm P}$ (for the poloidal component of the field) and $\xi_{\rm T}$ (for its toroidal part): 
\begin{equation}
\bB = \bnab \times \left[\bnab \times \left[\xi_{\rm P} \left(r, \theta\right){\widehat {\bf e}}_{r}\right] + \xi_{\rm T} \left(r, \theta\right){\widehat {\bf e}}_{r}\right]. 
\label{B_xi}
\end{equation}
Next, the vector potential $\mbf{A}$ is given by the relation $\bB = \bnab \times \mbf{A}$. Knowing that in the confined case the gauge choice is inconsequential, we can identify without further ado 
\begin{equation}
\mbf{A} = \bnab \times \left[\xi_{\rm P} \left(r, \theta\right){\widehat {\bf e}}_{r}\right] + \xi_{\rm T} \left(r, \theta\right){\widehat {\bf e}}_{r}.
\label{potA_xi}
\end{equation}
The magnetic stream functions are then projected on the spherical harmonics
\begin{eqnarray}
\xi_{\rm P}\left(r, \theta\right) &=& \sum_{\ell>0} \xi_0^\ell \left(r\right) Y_\ell ^0\left(\theta\right),\\
\xi_{\rm T}\left(r, \theta\right) &=& \sum_{\ell>0} \chi_0^\ell \left(r\right) Y_\ell ^0\left(\theta\right).
\end{eqnarray}
From now on, we use Einstein summation convention where $A^\ell \: B_\ell = \sum_\ell  A^\ell \: B_\ell$ and the vectorial spherical harmonics basis $\left({\bf R}_\ell ^0\left(\theta\right), {\bf S}_\ell ^0\left(\theta\right), {\bf T}_\ell ^0\left(\theta\right)\right)$ such that any axisymmetric vector field ${\vec u}(r,\theta)$ can be expanded as
\begin{equation}
{\vec u}(r,\theta)=u_{0}^{\ell}(r)\:{\bf R}_{\ell}^{0}\left(\theta\right)+v_{0}^{l}(r)\:{\bf S}_{\ell}^{0}\left(\theta\right)+w_{0}^{\ell}(r)\:{\bf T}_{\ell}^{0}\left(\theta\right),
\end{equation}
where the vectorial spherical harmonics ${\bf R}_{\ell}^{0}\left(\theta\right)$, ${\bf S}_{\ell}^{0}\left(\theta\right)$, and ${\bf T}_{\ell}^{0}\left(\theta \right)$ are defined by:
\begin{equation}
%\!\!\!\!\!
{\bf R}_{\ell}^{0}\left(\theta\right)=Y_{\ell}^{0}\left(\theta\right)\,\widehat{\bf e}_{r}\hbox{, }
{\bf S}_{\ell}^{0}\left(\theta\right)=\bnab_{\mathcal{S}}Y_{\ell}^{0}\left(\theta\right)\hbox{, }
%\end{equation}
%\begin{equation}
{\bf T}_{\ell}^{0}\left(\theta \right)=\bnab_{\mathcal{S}}\times{\bf R}_{\ell}^{0}\left(\theta \right)\!,
\end{equation}
the horizontal gradient being defined as
$
\bnab_{\mathcal{S}}=\etheta\: \partial_{\theta}
$. (cf. \cite{Rieutord:1987}). Since
\begin{equation}
\bnab \times \left( \xi_{\rm P}\,\widehat{\bf e}_{r}\right) = \bnab \times \left(\xi_0^\ell \: Y_\ell ^0\,\widehat{\bf e}_{r}\right)  = \bnab \times \left(\xi_0^\ell \: {\bf R}_\ell ^0 \right) = \frac{\xi_0^\ell}{r}{\bf T}_{\ell}^{0},
\end{equation}
we get from Eq. (\ref{potA_xi})
\begin{equation}
\mbf{A} = \chi_0^\ell \: {\bf R}_\ell ^0 + \frac{\xi_0^\ell}{r}  {\bf T}_{\ell}^{0}.
\label{potA_RST}
\end{equation}
On the other hand, we have
\begin{equation}
\bB = \frac{\ell \left(\ell +1\right) }{r^2} \xi_0^\ell \: {\bf R}_\ell ^0 
+ \frac{1}{r} \partial_r \xi_0^\ell \: {\bf S}_\ell ^0
+ \frac{\chi_0^\ell}{r}{\bf T}_{\ell}^{0}, 
\label{B_RST}
\end{equation}
from which we finally obtain the expression for the helicity
\begin{eqnarray}
\mathcal{H}&=& \int_0^{R_*} \!\!\int_\Omega\left\{\left[  \chi_0^\ell \: {\bf R}_\ell ^0 + \frac{\xi_0^\ell}{r}{\bf T}_{\ell}^{0}\right] \cdot\right. \nonumber \\
&&{\left.\left[\frac{{\ell'} \left({\ell'} +1\right) }{r^2} \xi_0^{\ell'} \: {\bf R}_{\ell'}^0 
+ \frac{1}{r} \partial_r \xi_0^{\ell'} \: {\bf S}_{\ell'} ^0
+ \frac{\chi_0^{\ell'}}{r}{\bf T}_{\ell'}^{0}\right]\right\}}{\rm d}\Omega\,r^2{\rm d}r.\qquad
\end{eqnarray}
At this point, we define a ``poloidal helicity'' defined by
\begin{equation} 
\mathcal{H}_{\rm P} = \int_{\mathcal V} \mbf{A}_{\rm P} \cdot  \bB_{\rm P} \: d{\mathcal V}
\end{equation}
and a ``toroidal helicity'' by
\begin{equation}
\mathcal{H}_{\rm T} = \int_{\mathcal V} A_\phi B_\phi \: d{\mathcal V}.
\end{equation}
Since $\left({\bf R}_\ell ^0 , {\bf S}_\ell ^0 , {\bf T}_\ell ^0 \right)$ constitutes an orthogonal basis,
\begin{equation}
\int_{\Omega}{\bf R}_{\ell}^{0}\cdot{\bf S}_{\ell}^{0}{\rm d}\Omega=\int_{\Omega}{\bf R}_{\ell}^{0}\cdot{\bf T}_{\ell}^{0}{\rm d}\Omega=\int_{\Omega}{\bf S}_{\ell}^{0}\cdot{\bf T}_{\ell}^{0}{\rm d}\Omega=0, 
\end{equation}
we get from the previous expression: 
\begin{eqnarray}
\mathcal{H} &=& \int_0^{R_*} \Bigg[ 
\frac{\ell^{'} \left(\ell^{'} +1\right) }{r^2}\: \chi_0^{\ell}\:\xi_0^{\ell^{'}} 
\int_\Omega {\bf R}_{\ell}^0\cdot{\bf R}_{\ell'}^0 \:d\Omega \nonumber \\
&+& \frac{1}{r^2}\:\xi_0^{\ell}\:\chi_0^{\ell'}
\int_\Omega {\bf T}_{\ell}^0\cdot{\bf T}_{\ell'}^0 \:d\Omega
\Bigg]\,r^2{\rm d}r
\label{heli_PT}
\end{eqnarray}
and we verify that 
\begin{equation}
\mathcal{H} = \mathcal{H}_{\rm P} + \mathcal{H}_{\rm T}.
\end{equation}
%en identifiant 
%$\mathcal{H}_{\rm P}= \mathcal{H}= \int_0^{R_*} \frac{ \ell \left(\ell +1\right) }{r^2} \xi_0^\ell \chi_0^{\ell'} \int_\Omega {\bf R}_{\ell}^0 {\bf R}_{\ell'}^0 \: r^2 d\Omega dr$  
%et 
%$\mathcal{H}_{\rm T} =  \int_0^{R_*}  \frac{1}{r^2}\chi_0^\ell \xi_0^{\ell'} \int_\Omega {\bf T}_{\ell}^0 {\bf T}_{\ell'}^0 \: r^2d\Omega dr$. 
From this expression (\ref{heli_PT}) we can draw two conclusions: 
\begin{enumerate}
\item {A magnetic field has to be mixed (both poloidal and toroidal) to be helical}; 
\item {The poloidal and the toroidal helicities are equal}%in the case where the radial field vanishes at the surface}
. This can be verified, exploiting the orthogonality relations
\begin{equation}
\int_\Omega {\bf R}_{\ell}^0\cdot{\bf R}_{\ell'}^0 \:d\Omega = \delta_{\ell,\ell'} 
%, \\
\end{equation}
and
\begin{equation}\int_\Omega {\bf T}_{\ell}^0\cdot{\bf T}_{\ell'}^0 \:d\Omega = \ell \left(\ell +1\right)\delta_{\ell,\ell'}, 
\end{equation}
$\delta_{\ell,\ell'}$ being the usual Kronecker symbol. Then, we get
\begin{equation}
\mathcal{H}_{\rm P} = \mathcal{H}_{\rm T} = \ell \left(\ell +1\right)  \int_0^{R_*} \xi_0^\ell\:  \chi_0^{\ell}\: dr = \mathcal{H}/2.
\label{EqualH}
\end{equation} 
\end{enumerate}

\subsection{Helicity {\it vs.} energy}
Now, we focus back again on the helicity expression in terms of the poloidal flux function $\Psi$. The equations (\ref{jPalphaBT}) and (\ref{jTalphaBP}) are rewritten as 
\begin{eqnarray}
\bB_{\rm P} &=&\frac{R}{\lambda_1^{0,i}} \bnab \times \bB_{\rm T} , 
\label{BProtBT}\\
\bB_{\rm T} &=&\frac{R}{\lambda_1^{0,i}} \bnab \times \bB_{\rm P} -\frac{R}{\lambda_1^{0,i}}\: \mu_0\: \beta_0\: {\overline\rho} \: r^2\: \sin \theta\: \ephi\nonumber \\
&= & \frac{R}{\lambda_1^{0,i}} \bnab \times \left[ \bB_{\rm P} -
\mu_0\: \beta_0\: {\overline\rho}\: r^2\: \cos \theta\: \er\right].
\label{BTrotBP}
\end{eqnarray}
We thus obtain the two vector potentials
\begin{eqnarray}
\mbf{A}_{\rm P} &=&\frac{R}{\lambda_1^{0,i}} \left( \bB_{\rm P} -
\mu_0\: \beta_0\: {\overline\rho}\: r^2\: \cos \theta\: \er\right) + \bnab  \Lambda_{\rm P},
\label{AToro}\\
\mbf{A}_{\rm T} &=&\frac{R}{\lambda_1^{0,i}} \bB_{\rm T} + \bnab  \Lambda_{\rm T}, 
\label{APolo}
\end{eqnarray}
$\Lambda_{\rm P}$ and $ \Lambda_{\rm T}$ being scalar gauge fields left free. When deriving the poloidal and toroidal helicities with the boundary condition $\bB\cdot{\widehat{\bf e}}_r = 0$ at the surface, these ones disappear after integration and we find : 
\begin{eqnarray}
{\mathcal H}_{\rm P} &=& \frac{2\: \mu_0 \: R}{\lambda_1^{0,i}} \int_{\mathcal{V}} \frac{\bB_{\rm P}^2}{2\: \mu_0} \:d{\mathcal V} -   \frac{\mu_0 \: R}{\lambda_1^{0,i}} \beta_0 \: \int_{\mathcal{V}}{\overline\rho}\: \Psi \: d{\mathcal V}, 
\label{HP_EP1}\\
{\mathcal H}_{\rm T} &=& \frac{2\: \mu_0 \: R}{\lambda_1^{0,i}} \int_{\mathcal{V}} \frac{\bB_{\rm T}^2}{2\: \mu_0} \:d{\mathcal V}.
\label{HT_ET1}
\end{eqnarray}
So, introducing the poloidal and toroidal magnetic energies ${U}_{\rm mag;P} = \int_{\mathcal{V}} \frac{\bB_{\rm P}^2}{2\: \mu_0} \:{\rm d}{\mathcal V}$ and ${U}_{\rm mag;T} = \int_{\mathcal{V}} \frac{\bB_{\rm T}^2}{2\: \mu_0} \:{\rm d}{\mathcal V}$ (respectively), we obtain: 
\begin{equation}
{\mathcal H}_{\rm P} = \frac{2\: \mu_0 \: R}{\lambda_1^{0,i}}\left({U}_{\rm mag;P} -   \frac{1}{2} \beta_0 M_{\Psi}\right), 
\label{HP_EP}
\end{equation}
where we identify $M_{\Psi}=\mathcal{I}_{I;1}=\int_{\mathcal{V}}{\overline\rho}\: \Psi \: d{\mathcal V}$, and
\begin{equation}
{\mathcal H}_{\rm T} = \frac{2\: \mu_0 \: R}{\lambda_1^{0,i}}{U}_{\rm mag;T}.
\label{HT_ET}
\end{equation}
Finally, adding these two last equations, we get the global relation between the helicity and the magnetic energy in the non force-free case
\begin{equation}
{\mathcal H}= \frac{2\: \mu_0 \: R}{\lambda_1^{0,i}} \left({U}_{\rm mag} -   \frac{1}{2}\: \beta_0 M_{\Psi}\right),
\label{H_E}
\end{equation}
where we recognize in the second term the non force-free contribution, which is the first invariant: the mass {enclosed} in magnetic flux surface.
\subsection{Helicity {\it vs.} topology}

{\it The $l>1$ latitudinal modes contributions}

As shown by \cite{Broderick:2008} for a set of modes $l$ ranging from 1 to 8 in the case of force-free solutions applied in an incompressible media, the first dipolar eigenvalue $\lambda_1^{0,1}$ corresponds to the minimum energy configuration. Furthermore, from the Eq. (\ref{H_E}), it arises directly that adding contributions from the higher multipolar components of the field (force-free) will result in adding a positive amount of magnetic energy to the total energy, and this one will not be the minimal state.\\
$\hbox{ }$\\
{\it Lowest energy radial mode}

We plotted in \fig \ref{Hsolap}a and \ref{Hsolap}b the poloidal, toroidal and total helicity in the case of the Sun and of the Ap star. 
It clearly stems from this figure that the poloidal and the toroidal helicities are equal (cf. Eq. \ref{EqualH}) for the eigenvalues given in {\sc Tab} \ref{eigenvals} (represented by red diamonds). Moreover, the energy of the poloidal component of the field ($U_{\rm mag;P}$) can be compared to the one correponding to the toroidal part ($U_{\rm mag;T}$) and we see that they are of the same order of magnitude.

In \fig \ref{HvsE}a and \ref{HvsE}b are represented the ratios $\mathcal{E}_{\rm mag}/\mathcal{H}$ for the poloidal, toroidal and global contributions, with and without the non force-free term, as a function of the parameter $\lambda_1^{0,i}$ respectively in the case of the Sun and of the Ap type star. The first dipolar eigenvalue $\lambda_1^{0,1}$ present the minimum energy compared with highest radial modes. It is thus the most probable configuration achieved after relaxation and from now on we focus on it.\\

\begin{figure}[h!]
\begin{center}
\includegraphics[width=0.49\textwidth]{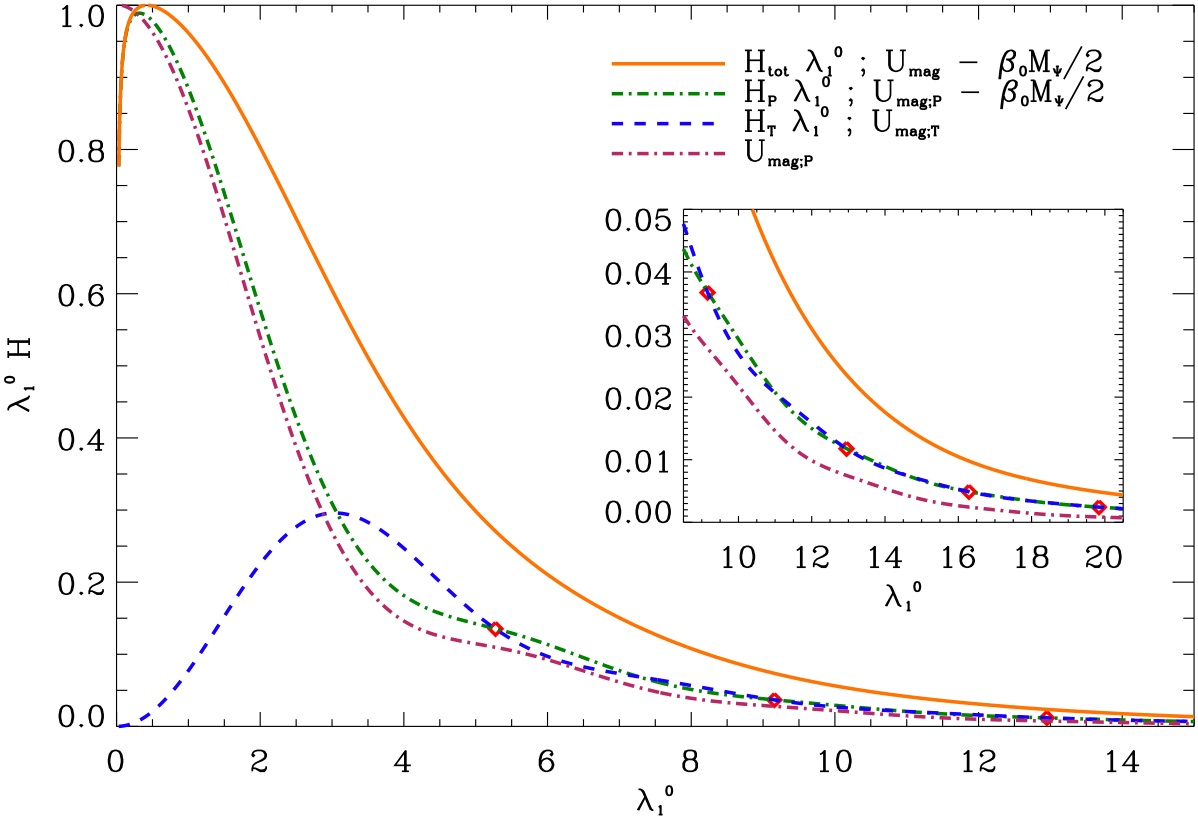}
\includegraphics[width=0.49\textwidth]{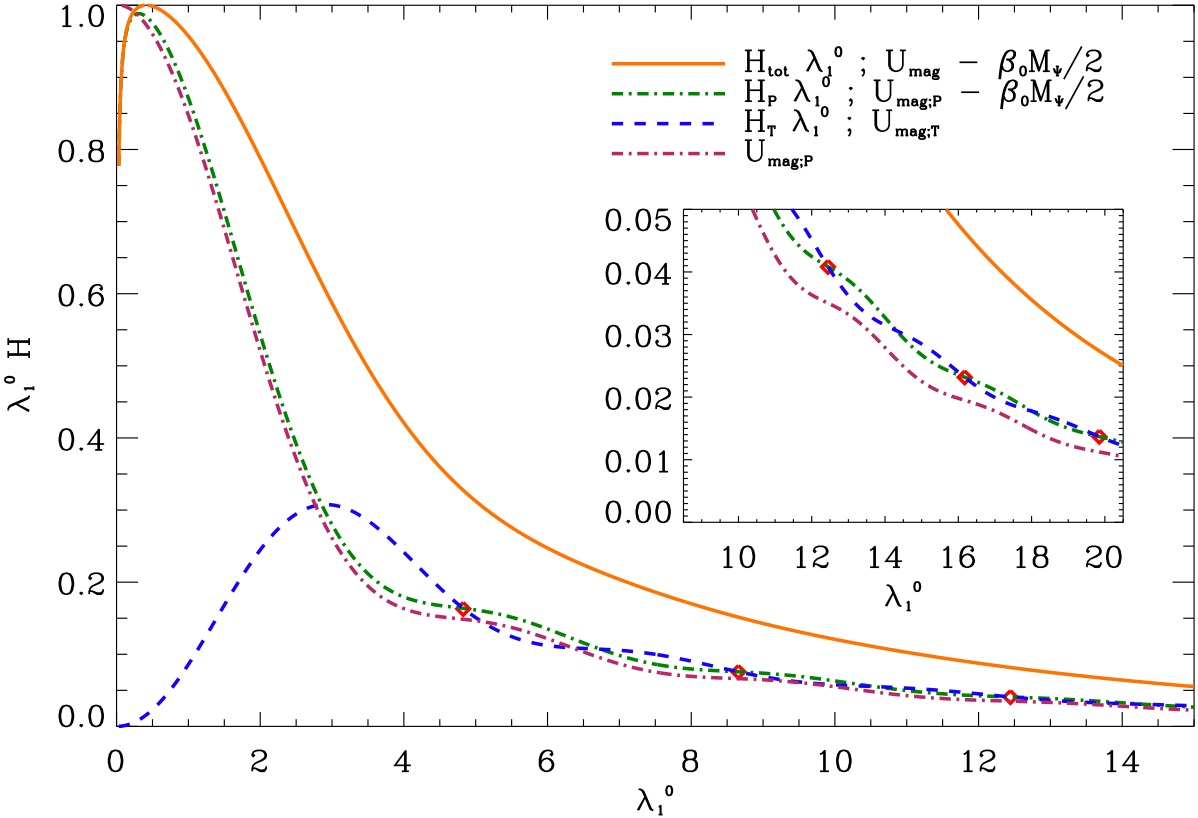}
\caption{Normalized total, poloidal, and toroidal helicities as a function of the eigenvalue ($\lambda_{1}^{0}$) in the case (a, top) of the young Sun and (b, bottom) of the studied Ap star. The red diamonds represent the eigenvalues ($\lambda_{1}^{0,i}$) given in  {\sc Tab} \ref{eigenvals} for which Eq. (\ref{EqualH}) is verified. Using Eq. (\ref{HT_ET}), we directly deduce $U_{\rm mag;T}$ while $U_{\rm mag;P}$ is given in purple.
\label{Hsolap}}
\vspace{-0.65cm}
\end{center}
\end{figure}
\begin{figure}[h!]
\begin{center}
\includegraphics[width=0.49\textwidth]{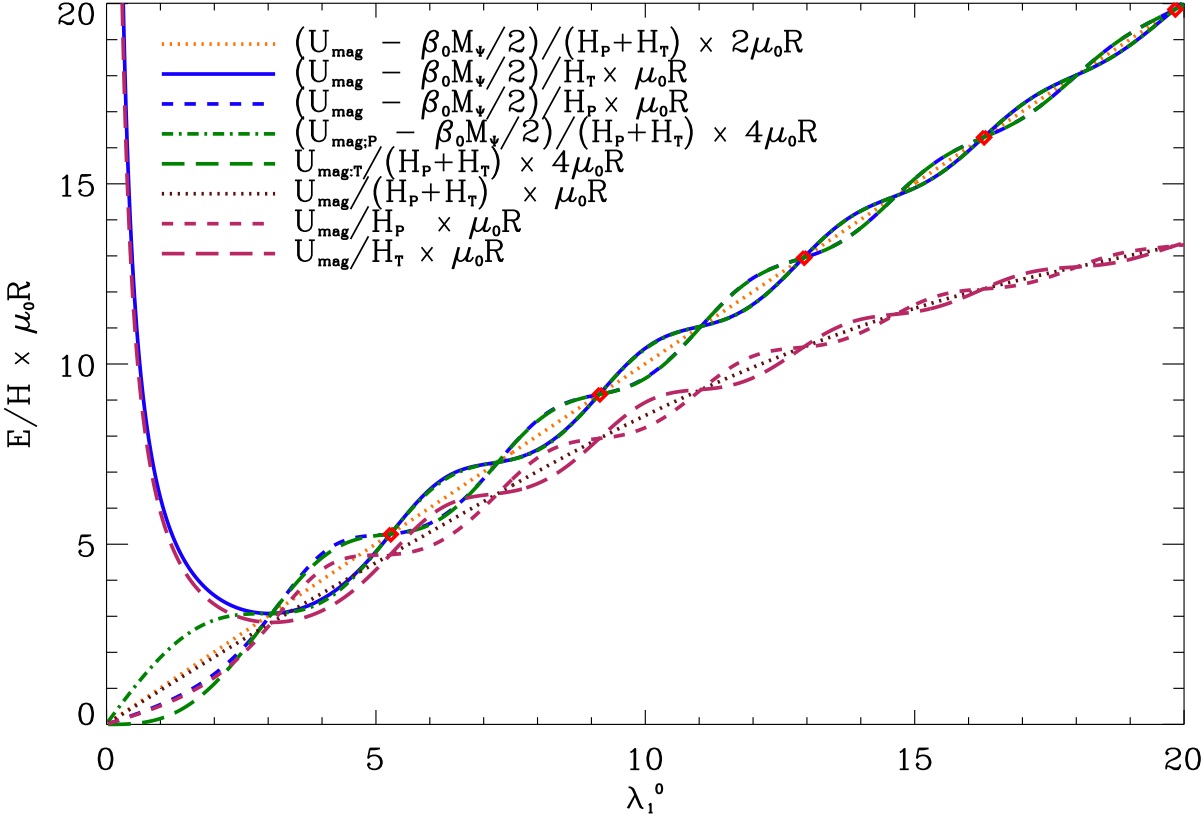}
\includegraphics[width=0.49\textwidth]{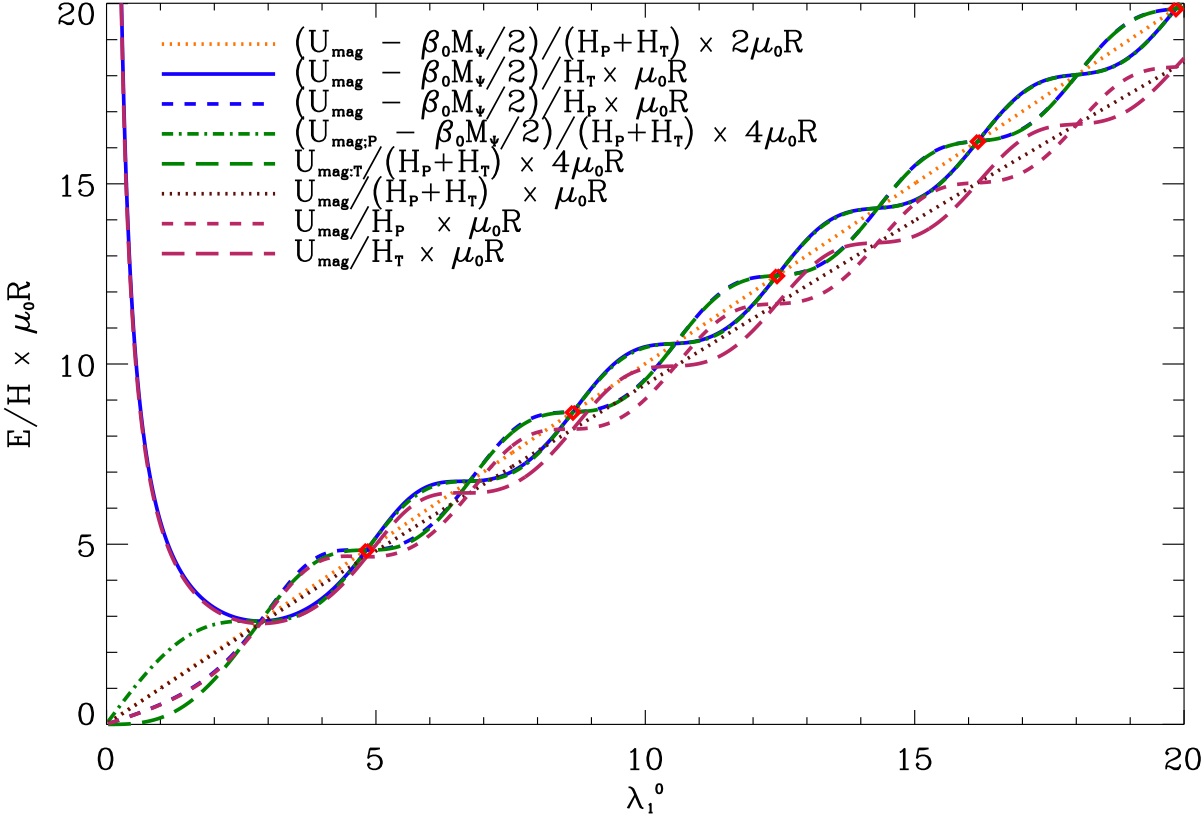}
\caption{Magnetic energy/helicity ratios for the total, poloidal, and toroidal contributions as a function of the eigenvalue ($\lambda_{1}^{0}$) with (first five curves) and without (last three curves) the non-force-free term in the case (a, top) of the young Sun and (b, bottom) of the studied Ap star. For the eigenvalues ($\lambda_{1}^{0,i}$) represented by the red diamonds (cf. {\sc Tab} \ref{eigenvals}), Eqs. (\ref{EqualH},\ref{H_E},\ref{HP_EP},\ref{HT_ET}) are simultaneously verified.
\label{HvsE}}
\vspace{-0.65cm}
\end{center}
\end{figure}
%
%****************************************************************************
%***************     Section : Discussion *********************************** 
%****************************************************************************
\section{Discussion}
\subsection{Stability criteria}
{First}, it is interesting to examine the ratio of the field's poloidal component amplitude with its toroidal one. Then, we define the anisotropy factor ($\gamma$) of the configuration\footnote{This can be inverted as\\
$\frac{B_{\rm T}}{B_{\rm P}}=\sqrt{\frac{1-\gamma}{1+\gamma}}\quad\hbox{and}\quad\frac{u_{\rm mag;T}}{u_{\rm mag;P}}=\frac{1-\gamma}{1+\gamma}$, \\
where the magnetic energy {densities} associated respectively with the poloidal field ($u_{\rm mag;P}=B_{\rm P}^2/ 2\mu_0$) 
%$E_{\rm P}=\frac{B_{\rm P}^2}{2\mu_0}$
and with the toroidal one ($u_{\rm mag;T}=B_{\rm T}^2/{2\mu_0}$)
%$E_{\rm T}=\frac{B_{\rm T}^2}{2\mu_0}$
have been introduced.} by
\begin{equation}
\gamma\left(r,\theta\right)=\frac{B_{\rm P}^2-B_{\rm T}^2}{B_{\rm P}^2+B_{\rm T}^2},\,\,\,\hbox{where}\,\,\,B_{\rm P}=\sqrt{B_{r}^{2}+B_{\theta}^2}\,.
\label{Anis}
\end{equation}
%\footnote{This can be inverted as\\
%$\frac{B_{\rm T}}{B_{\rm P}}=\sqrt{\frac{1-\gamma}{1+\gamma}}\quad\hbox{and}\quad\frac{E_{\rm T}}{E_{\rm P}}=\frac{1-\gamma}{1+\gamma}$, \\
%where the magnetic energy associated respectively with the poloidal field (
%$E_{\rm P}=B_{\rm P}^2/ 2\mu_0$
%$E_{\rm P}=\frac{B_{\rm P}^2}{2\mu_0}$
%) and with the toroidal one (
%$E_{\rm T}=B_{\rm T}^2/{2\mu_0}$
%$E_{\rm T}=\frac{B_{\rm T}^2}{2\mu_0}$
%) have been introduced.}

It runs between -1 when the field is completely toroidal to 1 when it is completely poloidal. In Figs. \ref{Sol1} \& \ref{Ap1}, it is shown for the first configurations obtained in the solar and in the Ap star cases. In both ones, the field is strongly toroidal $(\gamma\approx-1)$ in the center of the torus, which corresponds to the neutral point of the poloidal field (where we recall that $\partial_{r}\Psi=\partial_{\theta}\Psi=0$), while it is strongly poloidal $(\gamma\approx1)$ around the magnetic axis of the star where the toroidal field is weak. Between those two regimes, both components have comparable strengths where $\gamma\approx0$. Then, proposed configurations may be stable since poloidal and toroidal fields can stabilize each other respectively 
\citep{Wright:1973, Tayler:1980, Braithwaite:2009}. 
{The complete stability analysis 
following the analytical method given in \cite{Bernstein:1958} and using 3D numerical simulations will be achieved in a near future.}\\

\subsection{Comparison to numerical simulations}

{Next, let us compare in more details our analytical configuration to those obtained using numerical simulations (see \cite{Braithwaite:2004, Braithwaite:2006, Braithwaite:2008M}). 

Braithwaite and collaborators performed numerical magnetohydrodynamical simulations of the relaxation of an initially random magnetic field in a stably stratified star. Then, this initial magnetic field is always found to relax on the Alfv\'en time-scale into a stable magneto-hydrostatic equilibrium mixed configuration consisting of twisted flux tube(s). Two families are then identified: in the first one, the equilibria configurations are roughly axisymmetric with one flux tube forming a circle around the equator such as our configuration; in the second family, the relaxed fields are non-axisymmetric consisting of one or more flux tubes forming a complex structure with their axis lying at roughly constant depth under the surface of the star. Whether an axisymmetric or non-axisymmetric equilibrium forms depends on the initial condition chosen for the radial profile of the initial stochastic field strength $\vert{\mbf B}\vert\vert\propto{\overline\rho}^p$: a centrally concentrated one evolves into an axisymmetric equilibrium as in our configuration while a more spread-out field with a stronger connection to the atmosphere relaxes into a non-axisymmetric one. \cite{Braithwaite:2008M} indicates that, using an ideal-gas star modelled initially with a polytrope of index $n=3$, the threshold is $p\approx1/2$.

Moreover, as shown in Fig. 7 in \cite{Braithwaite:2008M}, a selective decay of the total helicity (${\mathcal H}$) and of the magnetic energy ($U_{\rm Mag}$) occurs during the initial relaxation with a stronger decrease of $U_{\rm Mag}$ than that of ${\mathcal H}$. This hierarchy well known in plasma physics (see for example \cite{Biskamp:1997} and \cite{Shaikhetal:2008}) justifies the variational method used to derive our configuration (\cite{MontgomeryPhillips:1989A}) while the introduction of ${\mathcal I}_{{\rm I};1}$ is justified by the non force-free character of the field in stellar interiors \citep{Reisenegger:2009} and by the stratification which inhibits the transport of flux and mass in the radial direction (see \S 3.2. and \cite{Braithwaite:2008M}).

Finally, note that our analytical configuration for which $U_{\rm mag;P}/U_{\rm mag}\approx0.45$ verifies the stability criterion derived by \cite{Braithwaite:2009} for axisymmetric configurations:
\begin{equation}
{\mathcal A}\frac{U_{\rm mag}}{U_{\rm grav}}<\frac{U_{\rm mag;P}}{U_{\rm mag}}\le0.8,
\end{equation}
where $U_{\rm grav}$ is the gravitational energy in the star and ${\mathcal A}$ a dimensionless factor whose value is of order $10$ in a main-sequence star and of order $10^3$ in a neutron star while we expect $U_{\rm mag}/U_{\rm grav}<10^{-6}$ in a realistic star (see for example \cite{Duezetal:2010}).

Our analytical solution is thus similar to the axisymmetric non force-free relaxed solutions family obtained by \cite{Braithwaite:2004} and \cite{Braithwaite:2006}.}\\

Those types of configurations can thus be relevant to model initial equilibrium conditions for evolutionary calculations involving large-scale fossil fields in stellar radiation zones. % as well as in degenerate objects such as white dwarfs or neutron stars. 
First, they can be used to initiate MHD rotational transport in dynamical stellar evolution codes where it is implemented 
\citep[cf.][]{Mathis:2005}
. There, axisymmetric transport equations have been derived to study the secular dynamics of the mean axisymmetric component of the magnetic field, the magnetic instabilities being treated using phenomenological prescriptions 
\citep{Spruit:1999, Spruit:2002,Maeder:2004} 
%(cf. Spruit 1999-2002, Maeder \& Meynet 2004) 
that have to be verified and improved by numerical experiments 
(\cite{Braithwaite:2006a} and subsequent works; \cite{Zahn:2007, Gellert:2008}). 
%(cf. Braithwaite 2006a and subsequent works; Zahn, Brun \& Mathis 2007; Gellert, R\"udiger \& Elstner 2008). 
On the other hand, those can also be used as initial conditions for large-scale numerical simulations of stellar radiation zones 
\citep{Garaud:2002, BrunZahn:2006}.\\
\subsection{Relaxed configurations and boundary conditions}

Let us now discuss the  boundary conditions we choosed. Since equilibrium states are known to minimize the energy/helicity ratio, we follow the procedure established by \cite{Chandrasekhar:1958} and \cite{Woltjer:1959b}
to constrain the arbitrary functions of the magnetohydrostatic equilibrium. This procedure, which minimizes the energy with respect to given invariants of the system (and in particular the helicity), assumes the following boundary condition $\bB\cdot \er=0$ that leads to an azimuthal current sheet due to the non-zero latitudinal field at the upper boundary ($B_{\theta}\left(R_{\rm sup},\theta\right)\ne0$). This is a potential source of instability and in the case of our configuration we have to evaluate its effect on the stability \citep[cf.][]{Bellan:2000}.

Next, in a stellar context, we have to allow open configurations as observed and thus to match the internal solution with an external multipolar one.  It remains then to
be seen whether the invariants are conserved, as they are in the confined case 
\citep{Dixon:1989}. 
%(cf. Dixon et al. 1989).
 
Finally, independently from the chosen type of configuration (confined or matched with a multipolar external field), we have to search solutions that allow the continuity of the magnetic field and of the associated currents at the boundaries to cancel the possible induced instabilities. This leads to an ill-posed problem which must be solved in a subtle way 
\citep[see][]{Monaghan:1976,Lyutikov:2009}. 
%(see Monaghan 1976 and Lyutikov 2009). 
In the present state of art, no solution has been derived that both minimizes the energy/helicity ratio and satisfies this type of surface boundary conditions. This will be the next step and it is out of the scope of the present paper.\\
%
%****************************************************************************
%***************     Section : Conclusion *********************************** 
%****************************************************************************
\section{Conclusion}
In the context of improving stellar models by taking into account in the most consistent way as possible dynamical processes such as rotation and magnetic field, we examine possible magnetic equilibrium configurations to model initial fossil fields. 

We generalize the pioneer work by Prendergast (1956) in deriving the barotropic magnetohydrostatic equilibrium states of realistic stellar interiors which are a first equilibrium family. These will then evolve due to other dynamical processes such as Ohmic diffusion, differential rotation, meridional circulation, and turbulence. Relaxed minimum energy equilibrium configurations are then obtained for a given conserved mass and helicity that correspond to the Taylor's relaxation states in the self-gravitating {non force-free case}.
%This equilibrium is then solved in the cases of confined fossil fields and of fossil fields matching at the surface with force-free potential ones. 
These are then applied to the internal radiation zone of the young Sun and to the radiative interior of an Ap star on the ZAMS. Mixed poloidal and toroidal magnetic configurations, potentially stable in stellar radiation zones, are obtained. 

Now, we have thus to study the stability of these magnetic topologies. Moreover, the case of general baroclinic equilibrium states have to be studied (Paper II).

These equilibrium configurations have then to be used as possible initial conditions for rotational transport processes in stellar radiative regions that will allow to study internal stellar MHD over secular time-scales. 

\acknowledgements{{We thank the referee for her/his remarks and suggestions that improved
and clarified the original manuscript.} We would like to thank S. Turck-Chi\`eze, A.-S. Brun, J.-P. Zahn, and {M. Rieutord} who kindly commented on the manuscript and suggested improvements and C. Neiner and G. Wade for valuable discussions on the subject. {This work was supported in part by the Programme National de Physique Stellaire (CNRS/INSU).}} 
%We thank the referee for her/his remarks that allowed to improve and to clarify the original manuscript. 

%Finally, it has to be examined if those correspond to minimum energy helicity relaxation states.
%
%****************************************************************************
%***************     Bibliography  ******************************************* 
%****************************************************************************
\bibliographystyle{aa} % style aa.bst
\bibliography{DM10arXiv} % your references Yourfile.bib

\end{document}